\documentclass[12pt,letterpaper]{article}
\usepackage[top=1in, bottom=1in, left=1in, right=1in]{geometry}

\usepackage{amsmath}
\allowdisplaybreaks

\usepackage{amssymb}

\usepackage{amsthm}

\usepackage[shortlabels]{enumitem}
\usepackage{footmisc}

\usepackage[title]{appendix}
\usepackage{setspace}

\usepackage{hyperref}
\hypersetup{
  hypertexnames=false,
  colorlinks=true,
  urlcolor=blue,
  linkcolor=blue,
  citecolor=blue}
\usepackage{natbib}
\DeclareMathOperator*{\argmax}{arg\,max}
\DeclareMathOperator*{\argmin}{arg\,min}

\DeclareMathOperator*{\rop}{relint}

\DeclareMathOperator*{\spa}{span}

\DeclareMathOperator*{\Fix}{Fix}

\newcounter{excounter}
\newtheorem{ex}[excounter]{Example}
\newtheorem*{exrevisited}{Example~\ref{ex-no-delay-multiplicity} revisited}
\newtheorem*{expatiencerevisited}{Example~\ref{ex-two-player-ties} revisited}
\newtheorem{lem}{Lemma}
\newtheorem{alem}{Lemma}[section]
\newtheorem{pr}{Theorem}
\newtheorem{prop}{Proposition}
\newtheorem{aprop}{Proposition}
\newtheorem{obs}{Observation}
\newtheorem{cor}{Corollary}
\theoremstyle{definition}
\newtheorem{definition}{Definition}

\newcounter{proofnumber}
\newcounter{proofstep}[proofnumber]

\AddToHook{env/proof/begin}{\stepcounter{proofnumber}}
\newcommand{\proofstep}[1]{\refstepcounter{proofstep}\textit{Step \theproofstep: #1}}
\newenvironment{assumption}[1]{\par\medskip\noindent\textbf{Assumption #1. }\itshape\ignorespaces}{\par\medskip}

\renewcommand{\theaprop}{A.\arabic{aprop}}

\title{Multiplicity and Uniqueness of Stationary Expected Payoffs in Coalitional Bargaining}
\author{Kirill S. Evdokimov\thanks{Department of Economics, Vanderbilt University, USA, \url{kirill.evdokimov@vanderbilt.edu}. OpenAI Codex was used for copy editing and proof checking. All errors are my own.}
}
\date{November, 2026}

\begin{document}

\maketitle

\begin{abstract}
    I study uniqueness of stationary expected payoffs in random-proposer coalitional bargaining with transferable utility. Proposer independence of coalition surplus does not guarantee uniqueness, even when the surplus is strictly monotone and every equilibrium is no-delay, meaning that passing is strictly suboptimal for every proposer. For $M^\natural$-concave characteristic functions, including non-monotone ones, I show that stationary expected payoffs are unique and continuous in the primitives, provided every stationary equilibrium is no-delay. To prove uniqueness, I identify an equitable grouping property of maximizing coalitions and show that $M^\natural$-concavity implies it at every no-delay candidate payoff vector. Using equitable grouping, I show that every no-delay equilibrium payoff vector is isolated and has positive local index. For parameter changes that keep every equilibrium no-delay, a uniform increase in a player's contribution to every coalition containing her strictly increases her discounted continuation payoff, while an increase in her discount factor weakly increases it. I apply these results to joint ventures with coordination costs and government formation with ministerial expertise.
\end{abstract}

\newpage
\section{Introduction}

Models of coalition formation are used in industrial organization, political economy, international trade, and public economics. See the survey by \cite{ray2015coalition}. Within the bargaining approach to coalition formation, a central question is whether the model yields a unique vector of equilibrium payoffs. Uniqueness makes comparative statics unambiguous without specifying an equilibrium-selection rule. It also provides a single-valued expected payoff map when bargaining is embedded in a larger model.\footnote{For example, \citet{yildirim2007proposal} studies competition for recognition, in which players expend resources to influence their probabilities of becoming proposer.} In bilateral bargaining, \cite{rubinstein1982perfect} established uniqueness of subgame perfect equilibrium payoffs. In multilateral environments, uniqueness of subgame perfect equilibrium payoffs can fail. Examples based on unanimity and simple-majority rules are discussed in \cite{sutton1986non} and \cite{baron1989bargaining}.

This paper studies stationary expected payoffs in a model of coalitional bargaining with random recognition and transferable utility. In every period, a randomly selected player either proposes an allocation of surplus within a feasible coalition or passes. If the proposal is accepted unanimously by the members of the proposed coalition, the game ends and the proposed allocation is implemented, with players outside the formed coalition receiving zero payoffs. The model extends the bargaining framework of \cite{baron1989bargaining} and is closely related to \cite{okada1996noncooperative,okada2011coalitional} and \cite{compte2010coalitional}.\footnote{For a survey of the literature building on \cite{baron1989bargaining}, see \cite{eraslan2019legislative}. In \cite{okada1996noncooperative,okada2011coalitional} bargaining can continue after a coalition forms and its members leave the game. We follow \cite{compte2010coalitional} and assume that bargaining ends when a coalition forms.}

The surplus generated by a coalition is proposer-independent and described by a characteristic function. The feasible family of coalitions is not assumed to be upward closed, and the characteristic function is not assumed to be monotone with respect to set inclusion. Therefore, we allow negative externalities. For example, adding a player to a feasible coalition can make the resulting coalition infeasible or reduce its total surplus.

This paper makes three contributions. First, Examples~\ref{ex-delay-multiplicity} and~\ref{ex-no-delay-multiplicity} show that proposer independence alone does not guarantee uniqueness. The first example uses delay to generate a reinforcing feedback between expected payoffs of potential coalition partners and the incentives to pass. Consider a player whose participation lowers coalition surplus. When she passes more often, the other players receive their expected payoffs more often, which raises their expected payoffs, making them more expensive to recruit. This example is similar to the multiplicity example in \cite{eraslan2017some} based on proposer-dependent surplus, although here the feedback arises with proposer-independent surplus. The second example shows, using three players, that multiplicity can also arise without delay, even when coalition surplus is strictly monotone.\footnote{\citet{yan2009multiplicity} also establishes no-delay multiplicity with proposer-independent surplus. We discuss her example in Section \ref{multiplicity}.} The positive results therefore focus on the no-delay domain and identify an additional local condition for uniqueness within that domain.\footnote{Note that equilibrium delay can be compatible with uniqueness; see Section \ref{conclusion}.} Appendix \ref{appendix-no-delay-conditions} gives sufficient conditions for no delay.

Second, Theorem~\ref{mnatural-unique} establishes uniqueness of equilibrium expected payoffs throughout the no-delay domain for $M^\natural$-concave characteristic functions, allowing the recognition probabilities and discount factors to differ across players.\footnote{\cite{fujishige2003note} show that $M^\natural$-concavity is equivalent to the gross substitutes condition of \citet{kelso1982job}. Gross substitutes and related properties have extensive applications in matching, Walrasian equilibrium \citep{gul1999walrasian}, and auction design \citep{gul2000english,ausubel2006efficient}. \citet{murota2016discrete} surveys discrete convex analysis and its applications in economics and game theory. \citet{paesleme2017gross} surveys gross substitutes and related properties from the algorithmic point of view.} The unique equilibrium payoff vector also depends continuously on the primitives. Corollary \ref{three-player-majority-unique} shows that in a game with three players and simple majority rule, equilibrium expected payoffs are unique for any characteristic function, provided every equilibrium is no-delay.

The proof uses fixed point index methods.\footnote{For a textbook treatment of fixed point theory in economics, see \cite{mclennan2018advanced}.} Equilibrium payoff vectors are characterized as fixed points of a correspondence with total index $+1$. This reduces uniqueness to showing that every equilibrium payoff vector is isolated and has positive local index. When every proposer has a unique optimal coalition, we show that these properties hold. Thus multiplicity necessarily involves an equilibrium at which some proposer is indifferent between multiple coalitions. To handle ties, we show that $M^\natural$-concavity implies an equitable grouping property of maximizing menus at every no-delay candidate payoff vector. This property restricts how tied coalitions differ within each proposer's menu and how recruitment patterns vary across proposers. We use this structure to deform the payoff correspondence locally to a constant map while preserving the equilibrium as its only nearby fixed point. This establishes isolation and local index $+1$.

Third, we apply these results to joint ventures with coordination costs and government formation with ministerial expertise, and establish comparative statics for productivity and patience. In the first application, coalition surplus combines firms' productive contributions with congestion and coordination costs across nested groups of firms, allowing surplus to be non-monotone. In the second, a coalition's surplus is determined by the optimal assignment of its members to cabinet positions. In both applications, the unique equilibrium payoff vector depends continuously on the underlying parameters within the no-delay domain.

The comparative statics hold for $M^\natural$-concave characteristic functions without further structural assumptions. We vary one primitive at a time over an interval on which every equilibrium is no-delay. A uniform increase in a player's contribution to every coalition containing her strictly raises her expected payoff, by no more than the surplus increase. Increasing her discount factor weakly raises her expected payoff, with equality possible only if the entire equilibrium payoff vector is unchanged. These comparisons allow maximizing menus and supporting proposal distributions to change.

Our approach builds on earlier applications of degree and fixed point index methods to coalitional bargaining. In a bargaining model with a general agreement space, \citet{kalandrakis2006regularity} characterizes pure-strategy stationary equilibria through expected payoff equations and inequalities ensuring optimal coalition choices.\footnote{In a stochastic bargaining model allowing mixed strategies, \citet{kalandrakis2014regular} establishes that there are generically finitely many equilibrium outcome distributions, each locally varying continuously with the parameters.} In a model with externalities in which bargaining continues after coalitions form, \citet{gomes2015multilateral} characterizes equilibria using a correspondence defined on coalition expected payoffs that incorporates optimal coalition choices. Both use degree or fixed point index methods to derive determinant conditions for global uniqueness. The condition in \cite{kalandrakis2006regularity} applies to oligarchic voting rules, and he verifies it for unanimity bargaining over a fixed surplus with linear utilities. \cite{gomes2015multilateral} verifies his condition for almost all three-player games in which the grand coalition is efficient. \citet{eraslan2013uniqueness} establish payoff uniqueness allowing proposer-dependent families of feasible coalitions, assuming every successful agreement divides the same fixed surplus.\footnote{\citet[Section~7]{eraslan2013uniqueness} ask whether uniqueness extends when different proposer--coalition pairs generate different pie sizes. Our examples show that uniqueness can fail even when coalition surplus is proposer-independent.} Their fixed point index argument uses a correspondence on proposer-specific inclusion probabilities. Our correspondence is instead defined directly on expected payoff space, analogous to \cite{gomes2015multilateral}. We establish positive local indices using equitable grouping and obtain primitive sufficient conditions for payoff uniqueness through $M^\natural$-concavity and no delay.

For bargaining over a fixed surplus, \cite{eraslan2002uniqueness} establishes uniqueness in the $q$-quota version of the \cite{baron1989bargaining} model with general recognition probabilities and discount factors. \citet{kalandrakis2015computation} develops a method for computing these payoffs that also recovers the uniqueness result. Theorem \ref{mnatural-unique} subsumes this uniqueness result because the unit-surplus $q$-quota model has an $M^\natural$-concave characteristic function and rules out delay.

For more general characteristic functions, \citet{yan2009uniqueness} gives two uniqueness results under uniform recognition probabilities and a common discount factor. The first establishes uniqueness for symmetric nondecreasing characteristic functions, without requiring concavity in coalition size. For convex (supermodular) characteristic functions, her second result establishes payoff equivalence among stationary equilibria in which no proposed coalition can be enlarged without reducing the proposer's surplus. Neither of these two results is subsumed by our uniqueness theorem for $M^\natural$-concave characteristic functions. However, Corollary~\ref{symmetric-concave-unique} establishes uniqueness for symmetric nondecreasing concave characteristic functions and allows recognition probabilities and discount factors to differ across players.

In another paper, \citet[Proposition~3(i)]{yan2009multiplicity} also establishes uniqueness with three players and simple majority rule for a monotone characteristic function, again under uniform recognition probabilities and a common discount factor. In contrast, Corollary~\ref{three-player-majority-unique} allows heterogeneous recognition probabilities and discount factors and non-monotone characteristic functions, provided every equilibrium is no-delay. Relatedly, \citet{okada2014stationary} presents a classification of stationary equilibria for sufficiently patient players in three-player games with superadditive characteristic functions, uniform recognition probabilities, and a common discount factor, and characterizes the corresponding limiting payoffs.

Under uniform recognition probabilities and a common discount factor, \cite{compte2010coalitional} give conditions for asymptotic efficiency of stationary equilibria and show that their limiting payoffs coincide with the coalitional Nash bargaining solution. Coalition-dependent surplus has also been studied under other bargaining protocols. \citet{ambrus2015continuous} establish uniqueness of Markov perfect equilibrium payoffs for monotone characteristic functions in a continuous-time model with random proposal opportunities and a finite deadline. \citet{battaglini2021coalition} studies a model in which a formateur selects a coalition and subsequently negotiates the allocation of its surplus, retaining the ability to switch coalitions during negotiations. He shows that a holdup problem can lead to the formation of inefficient coalitions.

Allowing coalition-dependent surplus makes it possible to study how productivity and voting rules affect payoff distribution and coalition composition. In the additive $q$-quota model of \citet{evdokimov2023equality}, coalition surplus is the sum of its members' productivities. With uniform recognition probabilities and a common discount factor, an increase in the quota weakly reduces inequality in expected payoffs, as measured by the Gini coefficient. \citet{casas2026coalition} study a $q$-quota model with productive and unproductive players, where coalition surplus consists of a fixed component $Y$ plus a common contribution $\eta$ from each productive member. When $Y/\eta$ is sufficiently large, equilibrium coalitions can include unproductive players while excluding productive ones, making coalition composition inefficient. In both environments, monotonicity and upward-closed feasibility rule out delay by Proposition \ref{monotone-no-delay}. The characteristic functions are $M^\natural$-concave, so Theorem \ref{mnatural-unique} recovers their payoff uniqueness results within a common framework.\footnote{Both papers allow feasible coalitions with zero surplus. Our theorem recovers their payoff uniqueness results when every feasible coalition has strictly positive surplus.}

The rest of the paper is organized as follows. Section~\ref{model} presents the model, characterizes equilibrium payoffs, and introduces the index argument. Section~\ref{multiplicity} presents the multiplicity examples. Section~\ref{no-delay-tie} introduces the no-delay restriction and the equitable grouping property, and shows that $M^\natural$-concavity implies this property. Section~\ref{uniqueness} establishes uniqueness and continuity of equilibrium payoffs. Section~\ref{verification} discusses applications, and Section~\ref{comparative-statics} derives comparative statics. Section~\ref{conclusion} concludes. The appendices provide sufficient conditions for no delay (Appendix \ref{appendix-no-delay-conditions}) and proofs (Appendix \ref{appendix}). The supplementary appendix derives the equilibrium payoffs in the multiplicity examples, establishes robustness of no-delay multiplicity, and characterizes no delay through a limited-harm condition.

\section{The model}\label{model}

The set of players is denoted by $N=\{1,\dots,n\}$, where $n\geq2$ is the number of players. Each player $i\in N$ is endowed with a vector of exogenous characteristics $(p_i,\delta_i)$, where $p_i>0$ is the probability that player $i$ becomes a proposer, $\sum_{i\in N}p_i=1$, and $\delta_i\in(0,1)$ is the discount factor of player $i$.

A \textit{coalition} is a subset $C\subseteq N$. An \textit{allocation} is a vector $x\in\mathbb R_+^n$, that is, $x_i\geq0$ for each $i\in N$. Each player $i\in N$ values only private consumption, and her utility equals her consumption. The set of \textit{feasible} coalitions is denoted by $\mathcal W$, with $\varnothing\not\in\mathcal W$. For each player $i\in N$, let $\mathcal W^i=\{C\in\mathcal W:i\in C\}$ denote the set of feasible coalitions that include player $i$, and assume that $\mathcal W^i$ is non-empty. Let $\chi:\mathcal W\to\mathbb R_{++}$ be the \textit{characteristic function} and note that $\chi(C)>0$ for all $C\in\mathcal W$. The characteristic function is not required to be monotone with respect to set inclusion, so the surplus generated by a coalition may decrease when players are added to it. Moreover, $\mathcal W$ is not required to be upward closed, so $C\in\mathcal W$ and $C\subset D$ do not necessarily imply $D\in\mathcal W$.

When acting as a proposer, player $i$ can either pass or propose to form any coalition $C\in\mathcal W^i$. Passing is associated with a special empty coalition $C_\varnothing=\varnothing$ that generates zero surplus, $\chi(C_\varnothing)=0$. An allocation $x$ is \textit{feasible} for coalition $C$ if $\sum_{j\in N}x_j\leq\chi(C)$ and $x_j=0$ for every $j\not\in C$. A pair $(C,x)$ is a \textit{proposal} for player $i$ if $C\in\mathcal W^i\cup\{C_\varnothing\}$ and $x$ is feasible for $C$. The set of all proposals that player $i$ can make is denoted by $\mathcal P_i$.

In each period $t=1,2,\dots$, player $i\in N$ becomes a proposer with probability $p_i$ and makes a proposal $(C,x)\in \mathcal P_i$. If the proposed coalition is $C_\varnothing$ (meaning that the proposer passes), the game moves to the next period. And if she makes a proposal $(C,x)$ with $C\in\mathcal W$, then the members of $C$ simultaneously vote to either accept the proposal or to reject it. If at least one player in $C$ rejects the proposal, the game moves to the next period. Otherwise, the allocation $x$ is implemented, each player $j\in N$ receives payoff $\delta_j^{t-1}x_j$ and the game ends. Each player receives a zero payoff if an agreement is never reached.

In this article, I focus on stationary strategies. A stationary mixed proposal strategy $\pi_i$ of player $i$ is a probability distribution over feasible proposals, that is, $\pi_i\in\Delta(\mathcal P_i)$, and a stationary voting strategy $\beta_i$ of player $i$ is a mapping $\beta_i:\mathbb R\to[0,1]$, where $\beta_i(x_i)$ is the probability that voter $i$ votes to accept the proposal that allocates $x_i$ to player $i$. In particular, the stationary proposal strategies are history-independent and the stationary voting strategies do not depend on the history of proposals, the proposer's identity, the proposed coalition, or the proposed shares to other players.

A stationary strategy profile $\sigma=(\pi,\beta)$ is a \textit{stationary subgame perfect equilibrium} if there are no profitable deviations given any history, any proposer $i$, and any proposal $(C,x)$. I restrict attention to stationary subgame perfect equilibria with weakly undominated voting strategies. For the remainder of the article, I will refer to these strategy profiles as \textit{equilibria}.

\subsection{Expected payoffs}

Any equilibrium strategy profile $\sigma$ induces a vector of expected payoffs $v(\sigma)\in\mathbb R_+^n$, where $v_i(\sigma)$ is the expected payoff of player $i$. Throughout the paper, $v(\sigma)$ denotes the payoff vector induced by a particular strategy profile $\sigma$. When no strategy profile is being fixed, I write $y$ for a generic expected payoff vector. Thus, $y$ is a candidate payoff vector until it is shown to satisfy the equilibrium equations. Let $\rho_i(\sigma)$ denote the distribution over coalitions induced by player $i$'s proposal strategy $\pi_i$, treating rejected proposals as passing. For any vector $z\in\mathbb R^n$ and subset of players $C\subseteq N$, write $z_C$ to denote $\sum_{i\in C}z_i$. Using this notation, for each $i\in N$, we can write:
\begin{align}
    \begin{split}\label{payoff}
        v_i(\sigma) &= \delta_ip_iS_i(v(\sigma)) + \delta_i\sum_{j\in N}p_j\sum_{C\in\mathcal C_j^i(v(\sigma))}\rho_j(\sigma)(C)v_i(\sigma), \\
        \rho_i(\sigma) &\in \Delta(\mathcal C_i(v(\sigma))),
    \end{split}\\
    \intertext{where}
    \begin{split}\label{problem}
        S_i(v(\sigma)) &= \max_{C\in\mathcal W^i\cup\{C_\varnothing\}}(\chi(C)-v_C(\sigma)), \\
        \mathcal C_i(v(\sigma)) &= \argmax_{C\in\mathcal W^i\cup\{C_\varnothing\}}(\chi(C)-v_C(\sigma)),
    \end{split}
\end{align}
and $\mathcal C_j^i(v)=\{C\in\mathcal C_j(v):i\in C\text{ or }C=C_\varnothing\}$. We call $S_i(v(\sigma))$ player $i$'s \textit{proposer surplus}, the greatest gain over her expected payoff that she can obtain when recognized. Her \textit{continuation probability} is $\sum_{j\in N}p_j\sum_{C\in\mathcal C_j^i(v(\sigma))}\rho_j(\sigma)(C)$, the probability that bargaining continues to the next period or a coalition containing $i$ forms.

The expected payoff $v_i(\sigma)$ is player $i$'s continuation payoff if the current proposal is rejected and the game moves to the next period, discounted to the current period. In an equilibrium, weakly undominated voting requires each player $i\in N$ to vote to accept any proposal $x$ with $x_i>v_i(\sigma)$ and to reject any proposal with $x_i<v_i(\sigma)$. When $x_i=v_i(\sigma)$, voter $i$ is indifferent. Without loss of generality, we assume that player $i$ votes to accept when indifferent,\footnote{If $S_i(v(\sigma))>0$, then in equilibrium an indifferent voter cannot reject proposer $i$'s proposal, since the proposer could profitably deviate by increasing the allocations of all potentially rejecting indifferent coalition members by sufficiently small amounts. Suppose instead that $S_i(v(\sigma))=0$. If $\chi(C)<v_C(\sigma)$ for every $C\in\mathcal W^i$, every feasible proposal with non-empty $C$ is rejected, so there is nothing to show. Otherwise, proposer $i$ is indifferent between passing and forming any coalition $C\in\mathcal W^i$ with $\chi(C)=v_C(\sigma)$. We can then modify strategies so that indifferent voters accept and proposer $i$ proposes each such coalition with its original probability of being formed, offering each member her expected payoff, and passes otherwise. Since rejection and passing have the same continuation effects, this preserves every player's expected payoff.} and that proposer $i$ passes instead of making a proposal that is rejected with probability one. A proposer will never offer a share to another player $i$ that is greater than $i$'s expected payoff $v_i(\sigma)$. Thus $v_i(\sigma)$ is the cost of player $i$'s participation in a coalition, and $v(\sigma)$ is the corresponding cost vector.

With these facts in mind, consider the components of the expected payoff $v_i(\sigma)$ for some player $i\in N$. With probability $p_i$, player $i$ becomes a proposer. If she proposes an empty coalition $C_\varnothing$, the game moves to the next period and $i$'s payoff equals $v_i(\sigma)$. If she forms a non-empty coalition $C$, then she allocates $v_j(\sigma)$ to each other coalition member $j\in C$ and nothing to other players. Combining these two cases, we can write the highest payoff that player $i$ can obtain when recognized as $v_i(\sigma)+\max_{C\in\mathcal W^i\cup\{C_\varnothing\}}(\chi(C)-v_C(\sigma))$, where for convenience it is assumed that player $i$ pays $v_i(\sigma)$ to herself when forming a non-empty coalition. If $S_i\bigl(v(\sigma)\bigr)>0$, player $i$ strictly prefers to form a non-empty coalition in $\mathcal C_i\bigl(v(\sigma)\bigr)$ and randomizes over these maximizers using the proposal distribution $\rho_i(\sigma)$. If $S_i\bigl(v(\sigma)\bigr)=0$, she is indifferent between proposing the empty coalition $C_\varnothing$ and any non-empty coalitions (if any) with $\chi(C)=v_C(\sigma)$, and randomizes over $\mathcal C_i\bigl(v(\sigma)\bigr)$. In either case, player $i$'s payoff (either as a share of surplus or as a continuation payoff) equals $v_i(\sigma)+S_i(v(\sigma))$. The expression in \eqref{payoff} accounts for the fact that player $i$ always includes herself when proposing a non-empty coalition, so $p_i\sum_{C\in\mathcal C_i^i(v(\sigma))}\rho_i(\sigma)(C)=p_i$ always holds.

When player $i$ is not a proposer, she votes to accept any proposal that allocates a share of surplus to her that is at least as large as her expected payoff $v_i(\sigma)$. Therefore, she receives $v_i(\sigma)$ when player $j\neq i$ is chosen to become a proposer, which happens with probability $p_j$, and either proposes an empty coalition or forms a non-empty coalition $C$ that includes player $i$. This means that player $i$ receives $v_i(\sigma)$ with a total probability $\sum_{j\neq i}p_j\sum_{C\in\mathcal C_j^i(v(\sigma))}\rho_j(\sigma)(C)$, giving us the remaining component of the summation in \eqref{payoff}.

\subsection{No-tie and tie payoff vectors}

Fix an equilibrium $\sigma$ and write $y=v(\sigma)$. For this payoff vector, let $K(y)=\{k\in N\mid |\mathcal C_k(y)|\ge 2\}$ be the set of proposers who are indifferent between multiple coalitions. For each $j\in N\setminus K(y)$, denote by $C_j(y)$ the unique coalition in $\mathcal C_j(y)$. Next, for each $i\in N$, let $C^i(y)=\{j\in N\setminus K(y)\mid i\in C_j(y)\text{ or }C_j(y)=C_\varnothing\}$, so that $p_{C^i(y)}$ is the total proposer probability of non-tied players whose selected coalition either contains $i$ or is empty. Define
\begin{equation}\label{total-inclusion}
    r^i(\sigma)=\sum_{k\in K(y)}p_k\sum_{\substack{C\in\mathcal C_k(y):\\ C\ni i\text{ or }C=C_\varnothing}}\rho_k(\sigma)(C),\qquad i\in N.
\end{equation}

If $K(y)=\varnothing$, then every proposer has a unique maximizing coalition and \eqref{payoff} becomes
\begin{equation}\label{payoff-interior}
    y_i = \delta_ip_iS_i(y)+\delta_ip_{C^i(y)}y_i,\qquad i\in N.
\end{equation}
An equilibrium $\sigma$ with $K(y)=\varnothing$ is called a \textit{no-tie} equilibrium, and the corresponding vector of expected payoffs $y=v(\sigma)$ is called a \textit{no-tie} payoff vector.

If $K(y)\neq\varnothing$, then tied proposers must satisfy indifference and mixing conditions. In particular, for each $k\in K(y)$ we have
\begin{equation}\label{indiff1}
    \chi(C)-y_C = \chi(D)-y_D, \quad \text{for all }C,D\in\mathcal C_k(y),
\end{equation}
and for each $i\in N$ we have
\begin{equation}\label{indiff2}
    y_i = \delta_ip_iS_i(y)+\delta_i\bigl(p_{C^i(y)}+r^i(\sigma)\bigr)y_i,
\end{equation}
where $r^i(\sigma)$ is given by \eqref{total-inclusion}. An equilibrium $\sigma$ with $K(y)\neq\varnothing$ is called a \textit{tie} equilibrium, and the corresponding vector of expected payoffs $y=v(\sigma)$ is called a \textit{tie} payoff vector.

\subsection{Characterization of expected payoffs}\label{characterization}

For each $i\in N$, let $Y_i=\{y_i\in\mathbb R\mid 0\leq y_i\leq\max_{C\in\mathcal W^i}\chi(C)\}$ be the set of possible expected payoffs for player $i$, and let $Y=\prod_{i\in N}Y_i$. This compact set contains every equilibrium payoff vector.

Extend the notation in \eqref{problem} to any candidate payoff vector $y\in Y$ by setting $S_i(y)=\max_{C\in\mathcal W^i\cup\{C_\varnothing\}}(\chi(C)-y_C)$ and $\mathcal C_i(y)=\argmax_{C\in\mathcal W^i\cup\{C_\varnothing\}}(\chi(C)-y_C)$. The notation $K(y)$, $C_j(y)$, and $C^i(y)$, and the tie/no-tie terminology, extend to these candidate vectors in the same way.

For $y\in Y$, define its \textit{type} by
\begin{equation}\label{tie-type-map}
    \mathcal A(y)=\Bigl(K(y),\ (\mathcal C_k(y))_{k\in K(y)},\ (C_j(y))_{j\notin K(y)}\Bigr).
\end{equation}
Define the \textit{aggregate inclusion polytope} at $y$ by
\begin{equation}\label{aggregate inclusion-polytope-at-y}
    R(y)=\left\{r\in[0,1]^n\mid r^i=\sum_{k\in K(y)}p_k\sum_{\substack{C\in\mathcal C_k(y):\\ C\ni i\text{ or }C=C_\varnothing}}\rho_k(C),\ \rho_k\in\Delta(\mathcal C_k(y)),\ k\in K(y)\right\}.
\end{equation}
If $S_i(y)>0$ for every $i\in N$, then
\begin{equation}\label{aggregate-inclusion-convex-hull}
    R(y)=\sum_{k\in K(y)}p_k\operatorname{co}\{\mathbf1_C\mid C\in\mathcal C_k(y)\},
\end{equation}
where $\operatorname{co}$ denotes convex hull and $\mathbf1_C\in\{0,1\}^n$ is the incidence vector of coalition $C$. Note that $R(y)=\{\mathbf 0\}$ when $K(y)=\varnothing$.

For $y\in Y$, define the payoff correspondence
\begin{equation}\label{correspondence}
    \mathcal T(y) = \left\{z\in\mathbb R^n\ \middle|\
        z_i=\delta_i p_iS_i(y)+\delta_i\bigl(p_{C^i(y)}+r^i\bigr)y_i,\quad r\in R(y)
    \right\},
\end{equation}
and write $\Fix(\mathcal T)=\{y\in Y:y\in\mathcal T(y)\}$ for its fixed point set.
$\mathcal T(y)$ collects the payoff updates obtained by taking $y$ as continuation payoffs and varying proposal distributions over maximizing coalitions.

The following proposition establishes equilibrium existence, characterizes equilibrium expected payoffs as fixed points of $\mathcal T$, and derives some further properties of $\mathcal T$ and $\Fix(\mathcal T)$. Its proof is standard and is given in Appendix \ref{appendix-payoff-correspondence}.

\begin{prop}\label{char-equiv}\label{payoff-correspondence}
    For $y\in Y$, we have $y\in\mathcal T(y)$ if and only if there exists an equilibrium $\sigma$ such that $v(\sigma)=y$. The correspondence $\mathcal T:Y\rightrightarrows Y$ is non-empty valued, compact valued, convex valued, and upper hemicontinuous. Moreover, the equilibrium payoff set $\Fix(\mathcal T)$ is non-empty and compact.
\end{prop}

\subsection{Fixed point index and additivity}

This section introduces the fixed point index of a correspondence, following \citet[Section 5]{eraslan2013uniqueness}. A correspondence $\Phi:\overline U\rightrightarrows Y$, where $U\subseteq Y$ is relatively open, is \textit{index admissible} if it is upper hemicontinuous with non-empty compact convex values and has no fixed point on the relative boundary of $U$. By \citet[Proposition 3]{eraslan2013uniqueness}, the fixed point index is the unique function $\Lambda_Y$ from the set of all index admissible correspondences to $\mathbb Z$ that satisfies Normalization, Additivity, and Continuity. We use the following properties:
\begin{enumerate}
    \item \textit{Normalization.} If $c:\overline U\to Y$ is a constant map whose value belongs to $U$, then $\Lambda_Y(c)=1$.
    \item \textit{Additivity.} If $\Phi:\overline U\rightrightarrows Y$ is index admissible and $U_1,\ldots,U_m\subseteq U$ are pairwise disjoint  relatively open sets such that $\operatorname{Fix}(\Phi)\subseteq\bigcup_{j=1}^{m}U_j$, then
    \[
        \Lambda_Y(\Phi)=\sum_{j=1}^{m}\Lambda_Y\left(\Phi|_{\overline U_j}\right).
    \]
    \item \textit{Homotopy invariance.} Let $H:\overline U\times[0,1]\rightrightarrows Y$ be jointly upper hemicontinuous and write $H_s=H(\cdot,s)$. If $H_s:\overline U\rightrightarrows Y$ is index admissible for every $s\in[0,1]$, then $\Lambda_Y(H_0)=\Lambda_Y(H_1)$.\footnote{Homotopy invariance follows from Continuity in \citet[Proposition 3]{eraslan2013uniqueness}. Because $H$ is jointly upper hemicontinuous and $\overline U\times Y$ is compact, Continuity implies that $s\mapsto \Lambda_Y(H_s)$ is locally constant and therefore constant on $[0,1]$.}
\end{enumerate}

Proposition 1 implies that the restriction of $\mathcal T$ to $\overline U$, denoted $\mathcal T|_{\overline U}:\overline U\rightrightarrows Y$, is index admissible whenever $U\subseteq Y$ is relatively open and $\mathcal T$ has no fixed point on the relative boundary of $U$. Let $\Lambda_Y(\mathcal T|_{\overline U})$ denote its \textit{local} index. If $U=Y$, let $\Lambda_Y(\mathcal T)$ denote its \textit{total} index.

\begin{lem}\label{degree-one}
    We have $\Lambda_{Y}(\mathcal T)=+1$.
\end{lem}

\begin{proof}
    Fix $c\in Y$ and consider the homotopy
    \[
        y\longmapsto(1-s)c+s\mathcal T(y), \qquad y\in Y,\quad s\in[0,1].
    \]
    By Proposition \ref{payoff-correspondence} and convexity of $Y$, this homotopy is jointly upper hemicontinuous with non-empty compact convex values in $Y$. Homotopy invariance and Normalization give $\Lambda_Y(\mathcal T)=\Lambda_Y(c)=+1$.
\end{proof}

\begin{lem}\label{local-degree-criterion}
    Suppose that every point of $\Fix(\mathcal T)$ is isolated relative to $Y$. Then $\Fix(\mathcal T)$ is finite. Moreover, for any collection $(U_y)_{y\in \Fix(\mathcal T)}$ of relatively open neighborhoods in $Y$ with pairwise disjoint closures such that $\overline U_y\cap \Fix(\mathcal T)=\{y\}$, we have
    \[
        1=\Lambda_{Y}(\mathcal T)=\sum_{y\in \Fix(\mathcal T)}\Lambda_{Y}(\mathcal T|_{\overline U_y}).
    \]
    If every local index in this sum is positive, then $\Fix(\mathcal T)$ is a singleton.
\end{lem}

\begin{proof}
    By Proposition \ref{payoff-correspondence}, $\Fix(\mathcal T)$ is non-empty and compact. Since every fixed point is isolated, $\Fix(\mathcal T)$ is finite. The neighborhoods in the statement contain every fixed point of $\mathcal T$, so Additivity gives
    \[
        \Lambda_{Y}(\mathcal T)=\sum_{y\in \Fix(\mathcal T)}\Lambda_{Y}(\mathcal T|_{\overline U_y}).
    \]
    By Lemma \ref{degree-one}, the left-hand side equals $+1$. If every term on the right-hand side is positive, each is an integer of at least one, and hence $|\Fix(\mathcal T)|=1$.
\end{proof}

By Lemma \ref{local-degree-criterion}, if every fixed point is isolated and has a positive local index then there is a unique fixed point. The next subsection shows that every no-tie fixed point is isolated and has local index $+1$. For tie fixed points, Section \ref{no-delay-tie} identifies additional conditions under which Section \ref{uniqueness} establishes isolation and a local index of $+1$.

\subsection{No-tie payoff vectors and local index}\label{deterministic-jacobian}

At a no-tie vector, each proposer $i$ has a unique maximizing coalition $C_i$. These coalitions remain the unique maximizers on a relatively open neighborhood $U$ of that vector in $Y$. Hence $\mathcal T$ is single-valued on $U$ and agrees with the affine map
\[
    t_i(y)=\delta_i p_iS_i(y)+\delta_i p_{C^i}y_i,\qquad i\in N.
\]
The map $y\mapsto y-t(y)$ has a constant Jacobian $J=I_n-D_y t$. Letting $d_i=1-\delta_i p_{C^i}>0$, $\gamma_i=\delta_i p_i>0$, and $W=\operatorname{diag}(d_1,\dots,d_n)$, we can write $J=W+\Gamma$, where $\Gamma_{ij}=\gamma_i\,\mathbb I\{j\in C_i\}$ and $\mathbb I\{\cdot\}$ denotes the indicator function.

\begin{lem}\label{deterministic-positive-det}\label{notie-local-degree}
    The matrix $J$ has strictly positive determinant at every no-tie vector. Moreover, every no-tie fixed point $y_0\in\Fix(\mathcal T)$ is isolated relative to $Y$ and, for every sufficiently small relatively open neighborhood $U$ of $y_0$ in $Y$, we have $\Lambda_Y(\mathcal T|_{\overline U})=+1$.
\end{lem}
\begin{proof}
    Fix a no-tie vector $y$, and write $C_i=C_i(y)$ and $S_i=\chi(C_i)-y_{C_i}$ for each $i\in N$.

    \proofstep{Order the players.}\label{step-notie-order} If $j\in C_i$ and $i\in C_j$, then $C_i=C_j$ and therefore $S_i=S_j$. Indeed, otherwise both coalitions would be feasible for both players, and uniqueness of maximizing coalitions would imply $\chi(C_i)-y_{C_i}>\chi(C_j)-y_{C_j}$ and $\chi(C_j)-y_{C_j}>\chi(C_i)-y_{C_i}$, a contradiction. Similarly, if $j\in C_i$ and $i\notin C_j$, then $C_i$ is feasible for player $j$, so uniqueness of $C_j$ implies $S_j>S_i$.

    Relabel the players so that $S_1\le\cdots\le S_n$, keeping players with the same chosen coalition consecutive.\footnote{Such relabeling applies the same permutation to the rows and columns of $J$ and therefore preserves its determinant.} Define $i\sim j$ if and only if $C_i=C_j$, and let $Q_1,\dots,Q_m$ be the corresponding equivalence classes labeled according to increasing $S_i$.

    \proofstep{Block upper-triangular structure.} If $C_i=C_\varnothing$, then $\Gamma_{ij}=0$ for all $j$, and moreover $\Gamma_{ji}=0$ for all $j\neq i$, because $i\in C_j$ for some $j\neq i$ would imply $S_i\ge S_j>0$, contradicting $S_i=0$. Now consider $i\in Q_k$ and $j\in Q_h$ with $k>h$ and both $C_i,C_j\neq C_\varnothing$. If $j\in C_i$, then Step \ref{step-notie-order} gives $S_j>S_i$, contradicting the ordering of the blocks. Hence $j\notin C_i$ and $\Gamma_{ij}=0$. Therefore $\Gamma$, and hence $J$, is block upper-triangular with diagonal blocks indexed by $Q_1,\dots,Q_m$.

    \proofstep{Determinant.}\label{step-notie-determinant} Because $J$ is block upper-triangular, $\det J$ is the product of the determinants of its diagonal blocks. If $C_i=C_\varnothing$ on a block $Q_k$, then $J_{Q_kQ_k}=W_{Q_kQ_k}$, with determinant $\prod_{i\in Q_k}d_i>0$. Otherwise, all players in $Q_k$ have the same chosen coalition, so $J_{Q_kQ_k}=W_{Q_kQ_k}+(\gamma_i)_{i\in Q_k}\mathbf 1_{\lvert Q_k\rvert}^\top$, where $(\gamma_i)_{i\in Q_k}$ is the vector of $\gamma_i$'s on that block and $\mathbf 1_{\lvert Q_k\rvert}$ is a vector of ones of length $\lvert Q_k\rvert$. By the matrix determinant lemma,\footnote{For any invertible $d\times d$ matrix $A$ and vectors $u,v\in\mathbb R^d$, $\det(A+uv^\top)=\det(A)(1+v^{\top}A^{-1}u)$.} $\det(J_{Q_kQ_k}) = \det(W_{Q_kQ_k})\bigl(1+\mathbf 1_{\lvert Q_k\rvert}^{\top}W_{Q_kQ_k}^{-1}(\gamma_i)_{i\in Q_k}\bigr) = \bigl(\prod_{i\in Q_k}d_i\bigr)\bigl(1+\sum_{i\in Q_k}\gamma_i/d_i\bigr) > 0$. Therefore $\det J>0$.

    \proofstep{Local index.}\label{step-notie-index} Now suppose the no-tie vector $y_0$ is a fixed point of $\mathcal T$. Choose a relatively open neighborhood $U$ of $y_0$ in $Y$ such that the maximizing coalitions remain unchanged on $\overline U$, so that $y_0$ is a fixed point of $t$ on $\overline U$. Consider the homotopy
    \[
        y\longmapsto(1-s)y_0+st(y),\qquad y\in\overline U,\quad s\in[0,1].
    \]
    This homotopy is jointly continuous and, since $Y$ is convex, takes values in $Y$. Because $t$ is affine with Jacobian $I_n-J$ and $y_0=t(y_0)$, we have
    \[
        y-\bigl((1-s)y_0+st(y)\bigr)=\bigl((1-s)I_n+sJ\bigr)(y-y_0).
    \]
    The matrix $(1-s)I_n+sJ$ has the same block structure as $J$, with $d_i$ replaced by $(1-s)+sd_i>0$ and $\gamma_i$ by $s\gamma_i\geq0$. The argument in Step \ref{step-notie-determinant} therefore gives $\det((1-s)I_n+sJ)>0$ for every $s\in[0,1]$. Thus, for every $s\in[0,1]$, the equation $y=(1-s)y_0+st(y)$ has the unique solution $y_0$ in $\overline U$. In particular, $y_0$ is an isolated fixed point of $\mathcal T$, and no map in the homotopy has a fixed point on the relative boundary of $U$. Homotopy invariance and Normalization give $\Lambda_Y(\mathcal T|_{\overline U})=+1$.
\end{proof}

Consequently, if equilibrium expected payoffs are not unique, at least one equilibrium payoff vector is a tie vector. Indeed, if every fixed point were no-tie, Lemma \ref{notie-local-degree} would make every fixed point isolated with local index $+1$, and Lemma \ref{local-degree-criterion} would imply uniqueness.

\section{Multiplicity of expected payoffs}\label{multiplicity}

The following examples show that expected payoffs need not be unique even when the characteristic function is independent of the proposer's identity. In the first example, multiplicity relies on delay.

\begin{ex}\label{ex-delay-multiplicity}
    Let $n=3$, $p_i=\frac13$, and $\delta_i=\frac{9}{10}$ for all $i\in N$. Let the collection of feasible coalitions be $\mathcal W=\{C\,:\,\lvert C\rvert\geq2\}$ and let the characteristic function be given by $\chi(C)=\min_{i\in C}a_i$ for all $C\in\mathcal W$, where each $a_i>0$ is an exogenous parameter. Lower values of $a_i$ can be interpreted as corresponding to a more harmful player. Further, let $a_1=\frac25$ and $a_2=a_3=1$.

    In this game, there are exactly three equilibrium expected payoff vectors. Let $\mu\in[0,1]$ be the probability that player 1 forms a feasible coalition, conditional on being recognized, and let $y(\mu)$ be the corresponding vector of expected payoffs. The three equilibrium payoff vectors are $y(0)=\left(0,\frac37,\frac37\right)$, $y\!\left(\frac13\right)=\left(0,\frac25,\frac25\right)$, and $y(1)=\left(\frac{6}{425},\frac6{17},\frac6{17}\right)$.
    
    In every equilibrium, players 2 and 3 propose the coalition $\{2,3\}$. When $\mu=0$, player 1 always proposes the empty coalition. When $\mu=1$, player 1 mixes uniformly between proposing $\{1,2\}$ and $\{1,3\}$. When $\mu=\frac13$, player 1 proposes the empty coalition with probability $\frac23$ and proposes $\{1,2\}$ and $\{1,3\}$ each with probability $\frac16$.\footnote{Supplementary Appendix~\ref{appendix-multiplicity-example} derives these payoff vectors.}
\end{ex}

The mechanism is related to the example in \citet[Section 3.3]{eraslan2017some}.\footnote{The models are different. In \cite{eraslan2017some}, surplus depends on the identity of the proposer. Specifically, in their three-player example, every coalition of size at least two generates surplus $.4$ when proposed by player 1 and surplus $1$ when proposed by players 2 or 3. Here, the characteristic function is proposer-independent and given by $\chi(C)=\min_{i\in C}a_i$.} For a given proposer surplus, a higher continuation probability raises a player's continuation payoff and therefore makes her a less attractive coalition partner. In this example, delay by player 1 creates a reinforcing feedback. When player 1 passes, players 2 and 3 both receive their continuation payoffs. In contrast, when player 1 forms a coalition only one of players 2 and 3 receives her continuation payoff. Therefore, when player 1 passes more often, the continuation payoffs of both players 2 and 3 increase, reinforcing player 1's incentive to pass.

The feedback mechanism in this example relies on delay caused by the non-monotonicity of the characteristic function. The next observation shows that there is no delay when $a_1$ is sufficiently high, even though adding player 1 to $\{2,3\}$ continues to reduce coalition surplus. We say that an equilibrium is \textit{no-delay}, or features immediate agreement, if passing is strictly suboptimal for every player.

\begin{obs}\label{negative-externality-no-delay}
    Fix the primitives of Example \ref{ex-delay-multiplicity}, except let $a_1\in(0,1)$ vary while $a_2=a_3=1$. If $a_1>\frac37$, then every equilibrium is no-delay.
\end{obs}

\begin{proof}
    Fix an equilibrium $\sigma$ and write $y=v(\sigma)$. When player 1 is recognized, players 2 and 3 together receive at most $y_2+y_3$. When player $i\in\{2,3\}$ is recognized, their combined payoff is at most $1$. If she forms a coalition, this follows because every feasible coalition generates surplus at most $1$. If she passes, $S_i(y)=0$, so equation \eqref{payoff} implies $y_i=0$, while the other player's continuation payoff is at most $1$. Averaging across recognition and discounting yields $y_2+y_3\leq\frac67<1$.

    Thus $S_i(y)\geq1-y_2-y_3\geq\frac17>0$ for $i\in\{2,3\}$, so players 2 and 3 strictly prefer to form a non-empty coalition to passing. Suppose that passing is optimal for player 1. Then $S_1(y)=0$, and equation \eqref{payoff} implies $y_1=0$. Optimality of passing requires $a_1-y_1-y_2\leq0$ and $a_1-y_1-y_3\leq0$. Hence $y_2,y_3\geq a_1>\frac37$, contradicting $y_2+y_3\leq\frac67$.
\end{proof}

Ruling out delay removes the feedback mechanism in Example~\ref{ex-delay-multiplicity}, but does not guarantee uniqueness. The next example has multiple equilibrium payoff vectors despite every equilibrium being no-delay and the characteristic function being strictly monotone.\footnote{\citet[Section~4]{yan2009multiplicity} constructs a 150-player example with a monotone, superadditive characteristic function, uniform recognition probabilities, and a common discount factor. In contrast, Example \ref{ex-no-delay-multiplicity} uses three players with heterogeneous recognition probabilities and discount factors.}

\begin{ex}\label{ex-no-delay-multiplicity}
    Let $n=3$, $p_1=\frac{15}{16}$, $p_2=p_3=\frac1{32}$, $\delta_1=\frac{16}{23}$, and $\delta_2=\delta_3=\frac{64}{65}$. Let the collection of feasible coalitions be $\mathcal W=2^N\setminus\{\varnothing\}$ and let the characteristic function be given by $\chi(\{1\})=1$, $\chi(\{2\})=\chi(\{3\})=2$, $\chi(C)=\frac52$ for every two-player coalition $C$, and $\chi(N)=5$. Thus every non-empty coalition is feasible, and coalition surplus is strictly increasing with inclusion.\footnote{Note that delay equilibria are ruled out because each singleton coalition is feasible. See Section \ref{sec-no-delay} for the argument.}

    In this game, there are exactly three equilibrium expected payoff vectors. Let $h\in[0,1]$ be the common probability that players 2 and 3 propose the grand coalition, conditional on being recognized, and let $y(h)$ be the corresponding vector of expected payoffs. The three equilibrium payoff vectors are $y(0)=\left(\frac{51}{23},\frac45,\frac45\right)$, $y\!\left(\frac12\right)=(2,1,1)$, and $y(1)=\left(\frac32,\frac75,\frac75\right)$.

    Player 1 always proposes the grand coalition. When $h=0$, players 2 and 3 always propose their own singletons. When $h=1$, they always propose the grand coalition. When $h=\frac12$, each mixes uniformly between proposing her own singleton and the grand coalition.\footnote{Supplementary Appendix~\ref{appendix-no-delay-multiplicity-payoffs} derives these payoff vectors and shows that, keeping $\mathcal W$ fixed, every sufficiently nearby $(\chi,p,\delta)$ admits at least three equilibrium payoff vectors, including when the symmetry between players 2 and 3 is broken.}
\end{ex}

Once again, the mechanism operates through continuation probabilities and continuation payoffs. Comparing the three equilibria, when players 2 and 3 form the grand coalition more often, the continuation probabilities of all players increase, putting an upward pressure on their continuation payoffs, which in turn puts a downward pressure on proposer surpluses. In this example, these counteracting forces lead to an increase in the continuation payoffs of players 2 and 3 and a decrease in the continuation payoff of player 1. Moreover, the fall in player 1's continuation payoff exceeds the rise in the continuation payoff of either player 2 or player 3. Hence the cost of recruiting players 1 and 3 falls for proposer 2, and the cost of recruiting players 1 and 2 falls for proposer 3, reinforcing their incentives to propose the grand coalition.

These examples show that ruling out delay removes one potential source of multiplicity of expected payoffs but does not guarantee their uniqueness.

\section{No delay and tie payoff vectors}\label{no-delay-tie}

This section first states the no-delay restriction motivated by Example \ref{ex-delay-multiplicity} and shows that it is robust to small perturbations of parameters. It then introduces the equitable grouping property and shows that $M^\natural$-concavity implies it. Section \ref{uniqueness} uses this property to establish isolation and a local index $+1$ at every no-delay equilibrium payoff vector.

Let $\Theta=\mathbb R_{++}^{\mathcal W}\times\rop\Delta^{n-1}\times (0,1)^n$ be the space of primitives $\theta=(\chi,p,\delta)$, where $\rop\Delta^{n-1}=\{p\in(0,1)^n:\sum_i p_i=1\}$. For a primitive $\theta\in\Theta$, objects from Section \ref{characterization} are evaluated at $\theta$, with this dependence written explicitly when useful. Let $\mathcal E(\theta)=\Fix(\mathcal T_\theta)$ denote the set of equilibrium expected payoff vectors under primitive $\theta$.

\subsection{No delay}\label{sec-no-delay}

A candidate payoff vector $y\in Y(\theta)$ is called no-delay if $S_i(y;\theta)>0$ for every $i\in N$, meaning that passing is strictly suboptimal for every proposer. Let $\Theta^{\mathrm{ND}}\subseteq\Theta$ denote the set of primitives for which every equilibrium payoff vector is no-delay.

\begin{assumption}{ND}
    The primitive satisfies $\theta\in\Theta^{\mathrm{ND}}$.
\end{assumption}

The next result shows that the no-delay restriction is robust to small perturbations of primitives. If nearby games had equilibria in which some player has zero proposer surplus, taking limits would yield such an equilibrium in the original game, contradicting Assumption ND. See Appendix \ref{appendix-no-delay} for the detailed proof.

\begin{lem}\label{open}
    The set $\Theta^{\mathrm{ND}}$ is open in $\Theta$.
\end{lem}

One sufficient condition for Assumption ND is that every singleton coalition is feasible. Indeed, if $S_i(y;\theta)=0$ for some player $i$ at an equilibrium payoff vector $y$, equation \eqref{payoff} implies $y_i=0$. Forming her feasible singleton would then yield strictly positive proposer surplus $\chi(\{i\})$, a contradiction. Appendix \ref{appendix-no-delay-conditions} gives further sufficient conditions for Assumption ND in both monotone environments and environments with negative externalities.

\subsection{Tie-difference space and equitable grouping}\label{tie-geometry}

For a no-delay candidate payoff vector $y\in Y$, let $\mathcal A=\mathcal A(y)$ be its type as defined in \eqref{tie-type-map}. Define the \textit{tie-difference space} as
\begin{equation}\label{type-tie-difference-space}
    L_{\mathcal A}=\spa\{\mathbf1_C-\mathbf1_D\mid C,D\in\mathcal C_k(y),\ k\in K(y)\}\subseteq\mathbb R^n,
\end{equation}
and let $L_{\mathcal A}^{\perp}\subseteq\mathbb R^n$ denote its orthogonal complement. By \eqref{indiff1}, for $C,D\in\mathcal C_k(y)$, we have $\chi(C)-y_C=\chi(D)-y_D$, or equivalently $(\mathbf1_C-\mathbf1_D)^\top y=\chi(C)-\chi(D)$. A change from $y$ to $y+h$ preserves this equality exactly when $(\mathbf1_C-\mathbf1_D)^\top h=0$. Because vectors $\mathbf1_C-\mathbf1_D$ span $L_{\mathcal A}$, its orthogonal complement $L_{\mathcal A}^{\perp}$ consists of all payoff movements $h$ that preserve all tie equalities within the maximizing menus at $y$.\footnote{Interpreting coalition indicators as bundles and continuation payoffs as prices, maximizing menus are demand sets. See \citet{baldwin2019understanding} for a geometric treatment of demand sets and indifference prices.} The next property restricts the structure of maximizing menus at a no-delay candidate payoff vector.

\begin{samepage}
\begin{definition}[Equitable grouping property]\label{def-group-structure}
A no-delay candidate payoff vector $y\in Y$ satisfies the \textit{equitable grouping property} if there exist pairwise disjoint non-empty groups $G_1,\dots,G_\ell\subseteq N$ such that:
\begin{enumerate}[(i),ref=\roman*]
    \item\label{maximizing-group-basis} the vectors $\mathbf1_{G_1},\dots,\mathbf1_{G_\ell}$ form a basis of $L_{\mathcal A(y)}^{\perp}$;
    \item\label{maximizing-recruitment-counts} for every $t,s\in\{1,\dots,\ell\}$, the number $|C\cap G_s|$ is the same for all $i\in G_t$ and $C\in\mathcal C_i(y)$.\footnote{Note that if $L_{\mathcal A(y)}^\perp=\{0\}$, both conditions hold with $\ell=0$.}
\end{enumerate}
\end{definition}
\end{samepage}

The equitable grouping property requires identifying disjoint groups of players that satisfy two properties. Part (i) requires that the tie-preserving payoff movements are exactly those that are constant within each group and zero outside the groups. Since a common payoff change within any group $G_s$ preserves every tie, the number $|C\cap G_s|$ must be constant over $C\in\mathcal C_i(y)$ for every proposer $i$. Indeed, for $h=\xi_s\mathbf1_{G_s}$ with $\xi_s\ne0$ and $C,D\in\mathcal C_i(y)$, we have $(\mathbf1_C-\mathbf1_D)^\top h=0$ and therefore $\mathbf1_C^\top\mathbf1_{G_s}=|C\cap G_s|=|D\cap G_s|=\mathbf1_D^\top\mathbf1_{G_s}$. Part (ii) additionally requires that these counts are constant across proposers in the same group.

It is worth looking at some examples of candidate and equilibrium payoff vectors satisfying or violating the equitable grouping property. The first example shows that the tie equilibrium payoff vector in Example \ref{ex-no-delay-multiplicity} violates it.
\begin{exrevisited}
    At the tie equilibrium payoff vector $y=(2,1,1)$, the maximizing menus are
    \[
        \mathcal C_1(y)=\{N\},\qquad
        \mathcal C_2(y)=\{\{2\},N\},\qquad
        \mathcal C_3(y)=\{\{3\},N\}.
    \]
    Consequently, $L_{\mathcal A(y)}=\spa\{(1,0,1),(1,1,0)\}$ and $L_{\mathcal A(y)}^\perp=\spa\{(-1,1,1)\}$. Since $L_{\mathcal A(y)}^\perp$ contains no nonzero group indicator, part (\ref{maximizing-group-basis}) fails.
\end{exrevisited}

The following example shows that the equitable grouping property can be violated because players in one group recruit different numbers of players from the same or another group.
\begin{ex}\label{ex-group-structure}
    Let $n=3$. Let the collection of feasible coalitions be $\mathcal W=2^N\setminus\{\varnothing\}$ and let the characteristic function be given by $\chi(\{1\})=5$, $\chi(\{2\})=\chi(\{3\})=3$, $\chi(C)=6$ for every two-player coalition $C$, and $\chi(N)=7$. At the candidate payoff vector $y=(2,2,2)$, the maximizing menus are
    \[
        \mathcal C_1(y)=\{\{1\}\},\qquad
        \mathcal C_2(y)=\{\{1,2\},\{2,3\}\},\qquad
        \mathcal C_3(y)=\{\{1,3\},\{2,3\}\}.
    \]
    We have $L_{\mathcal A(y)}=\spa\{(1,0,-1),(1,-1,0)\}$ and $L_{\mathcal A(y)}^\perp=\spa\{\mathbf1_N\}$. Thus part (\ref{maximizing-group-basis}) holds with the single group $G_1=N$. However, part (\ref{maximizing-recruitment-counts}) fails because proposer 1's maximizing coalition contains one member of this group, while those of proposers 2 and 3 contain two.
\end{ex}

To see that the equitable grouping property can hold, first consider the equilibrium payoff vector in the symmetric $q$-quota version of the Baron--Ferejohn model.\footnote{The equilibrium payoff vector is unique in this model; see \cite{eraslan2002uniqueness}.}
\begin{ex}\label{ex-bf-group-structure}
    Let $2\le q<n$, $p_i=1/n$, and $\delta_i=\delta\in(0,1)$ for every $i\in N$. Let the collection of feasible coalitions be $\mathcal W=\{C\subseteq N:|C|\ge q\}$ and let the characteristic function be given by $\chi(C)=1$ for every $C\in\mathcal W$. At the equilibrium payoff vector $y=\left(\delta/n,\dots,\delta/n\right)$, each proposer's maximizing menu consists of all $q$-player coalitions containing her. We have $L_{\mathcal A(y)}=\spa\{e_i-e_j:i,j\in N\}$ and $L_{\mathcal A(y)}^\perp=\spa\{\mathbf1_N\}$, where $e_i$ is the $i$th unit vector in $\mathbb R^n$. Thus both conditions hold with the single group $G_1=N$, since every maximizing coalition contains exactly $q$ members of this group.
\end{ex}

Finally, note that $L_{\mathcal A(y)}=\mathbb R^n$ implies $L_{\mathcal A(y)}^\perp=\{0\}$ and therefore the equitable grouping property holds trivially. The following example illustrates this possibility at an equilibrium payoff vector.
\begin{ex}\label{ex-two-player-ties}
    Let $n=2$, $p_1=p_2=\delta_1=\delta_2=\frac12$, $\mathcal W=2^N\setminus\{\varnothing\}$, $\chi(\{1\})=\chi(\{2\})=3$, and $\chi(\{1,2\})=4$. At $y=(1,1)$, the maximizing menus are
    \[
        \mathcal C_1(y)=\{\{1\},\{1,2\}\},\qquad
        \mathcal C_2(y)=\{\{2\},\{1,2\}\}.
    \]
    We have $L_{\mathcal A(y)}=\spa\{(1,0),(0,1)\}=\mathbb R^2$, so $L_{\mathcal A(y)}^\perp=\{0\}$. In fact, $y=(1,1)$ is an equilibrium payoff vector, at which both players form the grand coalition. 
\end{ex}

In the next subsection, we present a condition on the set of feasible coalitions $\mathcal W$ and the characteristic function $\chi$ that implies the equitable grouping property for all no-delay candidate payoff vectors.

\subsection{\texorpdfstring{$M^\natural$}{M-natural}-concave characteristic functions}\label{sec-mnatural}

We will now show that the equitable grouping property is implied by exchange properties of $\mathcal W$ and $\chi$. For any $C\subseteq N$ and $i,j\in N$, write $C-i=C\setminus\{i\}$ and $C+j=C\cup\{j\}$. The definition below follows \citet[Section 3.1]{murota2016discrete}. 

\begin{definition}[$M^\natural$-concavity]\label{def-mnatural-concavity}
A function $\chi:\mathcal W\to\mathbb R$ is \textit{$M^\natural$-concave} if, for any $C,D\in\mathcal W$ and $i\in C\setminus D$, either $C-i,D+i\in\mathcal W$ and
\begin{equation}\label{mnatural-transfer-inequality}
    \chi(C)+\chi(D)\leq\chi(C-i)+\chi(D+i),
\end{equation}
or there is $j\in D\setminus C$ such that $C-i+j,D+i-j\in\mathcal W$ and
\begin{equation}\label{mnatural-replacement-inequality}
    \chi(C)+\chi(D)\leq\chi(C-i+j)+\chi(D+i-j).
\end{equation}
\end{definition}

This condition requires that any player in $C\setminus D$ can be transferred from $C$ to $D$, possibly replacing a player in $D\setminus C$ who moves to $C$, while keeping both coalitions in $\mathcal W$ and without reducing their combined surplus. It therefore restricts both the set of feasible coalitions $\mathcal W$ and the values of the characteristic function $\chi$. Note that $M^\natural$-concavity does not imply that the characteristic function is monotone, as illustrated in Section \ref{sec-laminar}.

Subtracting $y_C$ from $\chi(C)$ for each coalition $C\in\mathcal W$ preserves $M^\natural$-concavity because moving players to $D$ or exchanging players with $D$ leaves the sum $y_C+y_D$ unchanged. The following lemma derives the equitable grouping property from $M^\natural$-concavity.

\begin{lem}\label{mnatural-tie-structure}
    Fix $\theta\in\Theta$ and suppose that $\chi$ is $M^\natural$-concave on $\mathcal W$. Every no-delay candidate payoff vector $y\in Y(\theta)$ satisfies the equitable grouping property.
\end{lem}

The proof is in Appendix \ref{appendix-mnatural-tie-structure}. It shows that, under $M^\natural$-concavity, the tie-difference space $L_{\mathcal A(y)}$ defined in \eqref{type-tie-difference-space} can be generated by single-player additions, removals, and replacements within maximizing menus.\footnote{These elementary changes correspond to the strong-substitutes demand directions in \citet[Definition~3.9 and Proposition~3.10]{baldwin2019understanding}.} Using this representation, we define the groups whose indicators span $L_{\mathcal A(y)}^\perp$, giving part \eqref{maximizing-group-basis}. For part \eqref{maximizing-recruitment-counts}, the exchange inequalities imply that players who can be substituted for one another within a maximizing menu have equal proposer surpluses. Players within each group $G_t$ are connected by chains of such replacements, so their proposer surpluses coincide. By part \eqref{maximizing-group-basis}, a sufficiently small common increase $\xi_s$ in the payoffs of players in any group $G_s$ preserves all maximizing menus, so these surplus equalities continue to hold. For each proposer $i\in G_t$, her proposer surplus falls by $\xi_s$ times the number of members of $G_s$ in her maximizing coalitions. These numbers must therefore agree across proposers in the same group $G_t$, giving part \eqref{maximizing-recruitment-counts}.

The converse of Lemma \ref{mnatural-tie-structure} does not hold. The main reason is that the equitable grouping property only restricts maximizing menus, while $M^\natural$-concavity restricts all feasible coalitions. For instance, the exchange inequalities can fail because of coalitions that never maximize the proposer surplus on $Y$. The following example illustrates this point.

\begin{ex}\label{ex-uniform-grouping-not-mnatural}
    Let $n=3$ and $\mathcal W=2^N\setminus\{\varnothing\}$. Let the characteristic function be given by $\chi(\{i\})=3$ for every $i\in N$, $\chi(\{1,2\})=4$, and $\chi(\{1,3\})=\chi(\{2,3\})=\chi(N)=1$. At every $y\in Y$, each coalition $C$ with $\chi(C)=1$ yields strictly less proposer surplus than the proposer's singleton, since $1-y_C<3-y_i$ for every $i\in C$. Thus, at every no-delay candidate payoff vector, we have
    \[
        \mathcal C_1(y)\subseteq\{\{1\},\{1,2\}\},\qquad
        \mathcal C_2(y)\subseteq\{\{2\},\{1,2\}\},\qquad
        \mathcal C_3(y)=\{\{3\}\}.
    \]
    The only possible ties give the differences $e_2$ and $e_1$, respectively. Hence $L_{\mathcal A(y)}$ is spanned by a subset of $\{e_1,e_2\}$. The remaining unit vectors in $\{e_1,e_2,e_3\}$ form a basis of its orthogonal complement $L_{\mathcal A(y)}^\perp$. Taking the corresponding players as singleton groups, part (i) holds by construction. For example, if $L_{\mathcal A(y)}=\spa\{e_1\}$, then $L_{\mathcal A(y)}^\perp=\spa\{e_2,e_3\}$ and the groups are $G_1=\{2\}$ and $G_2=\{3\}$. For part (ii), note that any tie in the maximizing menus above changes the membership of just one player, whose unit vector therefore belongs to $L_{\mathcal A(y)}$. This player lies outside the groups, so $|C\cap G_s|$ is constant over $C\in\mathcal C_i(y)$ for every $i$ and $s$. However, $\chi$ is not $M^\natural$-concave. Taking $C=\{1,2\}$, $D=\{3\}$, and $i=1$, we have
    \[
        \chi(\{1,2\})+\chi(\{3\})=7>4
        =\chi(\{2\})+\chi(\{1,3\})
        =\chi(\{2,3\})+\chi(\{1\}).
    \]
    Thus both the transfer inequality \eqref{mnatural-transfer-inequality} and the only possible replacement inequality \eqref{mnatural-replacement-inequality} fail.
\end{ex}

\section{Uniqueness under \texorpdfstring{$M^\natural$}{M-natural}-concavity}\label{uniqueness}

Under $M^\natural$-concavity, the next lemma establishes isolation and a local index of $+1$ at no-delay equilibrium payoff vectors with ties, extending Lemma \ref{notie-local-degree}.

\begin{lem}\label{structural-local-index}
    Fix $\theta\in\Theta$ and suppose that $\chi$ is $M^\natural$-concave on $\mathcal W$. Every no-delay equilibrium payoff vector $y_0\in\mathcal E(\theta)$ is isolated and for every sufficiently small open neighborhood $U_{y_0}$ of $y_0$ in $\mathbb R^n$, we have $\Lambda_{Y(\theta)}(\mathcal T_\theta|_{\overline U_{y_0}})=+1$.
\end{lem}

The proof, presented in Appendix \ref{appendix-structural-local-index}, uses the equitable grouping property established in Lemma \ref{mnatural-tie-structure}. For isolation, compare $y_0$ with a sufficiently nearby equilibrium payoff vector $y$, so that $\mathcal C_i(y)\subseteq\mathcal C_i(y_0)$ for every proposer $i$. The proof shows that there exist proposal distributions over these surviving menus that reproduce the inclusion probabilities supporting $y_0$. Part \eqref{maximizing-group-basis} of equitable grouping at $y$ then implies that payoff differences $y_i-y_{0,i}$ are common within each group and zero outside the groups, while inclusion-probability differences sum to zero within each group. Part \eqref{maximizing-recruitment-counts} makes proposer-surplus differences common across proposers in each group, because they recruit the same number of players from every group. Subtracting the payoff equations at $y_0$ from those at $y$, rearranging, and summing within groups eliminates the inclusion-probability differences. Expressing proposer-surplus differences in terms of the group payoff differences using the common recruitment counts then gives a homogeneous system for the group payoff differences. An adaptation of the ordering and determinant argument in Lemma \ref{notie-local-degree} shows that this system is nonsingular, implying $y=y_0$.

Lemma \ref{local-degree-criterion} and Lemma \ref{structural-local-index} give our main result.
\begin{pr}\label{mnatural-unique}
    Fix $\theta\in\Theta^{\mathrm{ND}}$ and suppose that $\chi$ is $M^\natural$-concave on $\mathcal W$. Then the equilibrium expected payoff vector is unique.
\end{pr}

\begin{proof}
    Since $\theta\in\Theta^{\mathrm{ND}}$, every equilibrium payoff vector is no-delay. Lemma \ref{structural-local-index} therefore shows that every fixed point of $\mathcal T_\theta$ is isolated and has local index $+1$. Lemma \ref{local-degree-criterion} implies that $\mathcal E(\theta)$ is a singleton.
\end{proof}
The characteristic function in Example~\ref{ex-delay-multiplicity} is $M^\natural$-concave for every choice of the positive parameters $(a_1,a_2,a_3)$. Thus, keeping the other primitives of Example~\ref{ex-delay-multiplicity} fixed and allowing $a_1$ to vary, Observation~\ref{negative-externality-no-delay} and Theorem~\ref{mnatural-unique} give a unique equilibrium payoff vector whenever $a_1\in(3/7,1)$.

More generally, with three players and simple majority rule, no delay is sufficient for uniqueness because every characteristic function on this feasible family is $M^\natural$-concave. Indeed, for two distinct pairs $\{i,j\}$ and $\{j,k\}$, exchanging $i$ and $k$ gives the same pair of coalitions. For the grand coalition $\{i,j,k\}$ and a pair $\{i,j\}$, transferring $k$ again gives the same pair of coalitions. The exchange inequalities in Definition \ref{def-mnatural-concavity} therefore hold with equality and Theorem~\ref{mnatural-unique} gives the following corollary.\footnote{An analogous argument applies for any $n>3$ with $\mathcal W=\{C\subseteq N:|C|\geq n-1\}$.}
\begin{cor}\label{three-player-majority-unique}
    Suppose that $n=3$ and $\mathcal W=\{C\subseteq N:|C|\geq2\}$. For every $\theta\in\Theta^{\mathrm{ND}}$, the equilibrium expected payoff vector is unique.
\end{cor}

\citet[Proposition~3(i)]{yan2009multiplicity} establishes uniqueness of equilibrium expected payoffs with three players and simple majority rule when the characteristic function is monotone, recognition probabilities are uniform, and discount factors are common.\footnote{She also allows singleton coalitions, but they have zero surplus and are strictly suboptimal in every equilibrium when the grand coalition has positive surplus. In contrast, we assume that $\chi(C)>0$ for every feasible $C$.} Corollary~\ref{three-player-majority-unique} allows heterogeneous recognition probabilities and discount factors and non-monotone characteristic functions, provided Assumption ND holds.

Example~\ref{ex-no-delay-multiplicity} shows that no-delay multiplicity is possible when singleton coalitions are also feasible. The characteristic function used in that example is not $M^\natural$-concave.\footnote{Of course, this also follows from Lemma \ref{mnatural-tie-structure} and Example \ref{ex-no-delay-multiplicity} revisited.} To see this, take $C=N$, $D=\{1\}$, and $i=2$. Then $D\setminus C=\varnothing$, so replacement in \eqref{mnatural-replacement-inequality} is impossible. In turn, the transfer inequality \eqref{mnatural-transfer-inequality} fails because $\chi(\{1\})+\chi(N)=6>5=\chi(\{1,2\})+\chi(\{1,3\})$. 

Symmetric characteristic functions that are concave in coalition size are $M^\natural$-concave on $\mathcal W$ generated by quota rules. To see this, write $\chi(C)=f(|C|)$ with $f$ concave, and take $C,D\in\mathcal W$ and $i\in C\setminus D$. If $D\setminus C$ is non-empty, replacing $i$ with any $j\in D\setminus C$ preserves both coalition sizes, so \eqref{mnatural-replacement-inequality} holds with equality. If $D\setminus C$ is empty, then $D$ is a proper subset of $C$, so transferring $i$ from $C$ to $D$ preserves feasibility, and concavity gives $f(|C|)+f(|D|)\leq f(|C|-1)+f(|D|+1)$, so \eqref{mnatural-transfer-inequality} holds. If $f$ is also nondecreasing, Proposition~\ref{monotone-no-delay} implies no delay and Theorem~\ref{mnatural-unique} gives the following corollary.
\begin{cor}\label{symmetric-concave-unique}
    Fix $q\in\{1,\ldots,n\}$ and let $\mathcal W=\{C\subseteq N:|C|\geq q\}$. Suppose that $\chi(C)=f(|C|)$ for a nondecreasing concave function $f:\{q,\ldots,n\}\to\mathbb R_{++}$. Then the equilibrium expected payoff vector is unique for every $p\in\rop\Delta^{n-1}$ and $\delta\in(0,1)^n$.
\end{cor}

This result complements \citet[Proposition~3]{yan2009uniqueness}, who establishes uniqueness for symmetric nondecreasing characteristic functions under uniform recognition probabilities and a common discount factor, without requiring concavity in coalition size. Under concavity, Corollary \ref{symmetric-concave-unique} allows recognition probabilities and discount factors to differ across players.

The proof of Lemma~\ref{open} establishes that $\theta\mapsto\mathcal E(\theta)$ has a closed graph and that equilibrium payoff vectors belong to a common compact set for all primitives in a neighborhood of any $\theta\in\Theta$. These properties imply that $\theta\mapsto\mathcal E(\theta)$ is upper hemicontinuous. Together with uniqueness in Theorem~\ref{mnatural-unique}, this gives the following corollary.
\begin{cor}\label{equilibrium-payoff-continuity}
    On the set of primitives $\theta\in\Theta^{\mathrm{ND}}$ for which $\chi$ is $M^\natural$-concave on $\mathcal W$, the unique equilibrium expected payoff vector depends continuously on $\theta$.
\end{cor}

Finally, under Assumption ND, uniqueness of equilibrium expected payoffs also implies uniqueness of continuation probabilities. Let $y$ denote the unique equilibrium expected payoff vector. Since $y_i>0$ for every $i\in N$, equation \eqref{payoff} uniquely determines each player's continuation probability. Under no delay, these are the probabilities of being included in the formed coalition, so their sum gives the expected coalition size, which is therefore unique. Because the agreement is immediate and the coalition surplus is allocated fully, the expected surplus equals $\sum_{i\in N}y_i/\delta_i$ and is also unique.

\section{Applications}\label{verification}\label{mnatural-examples}

Many classes of $M^\natural$-concave functions can be used to model coalition surplus; see \citet{murota2016discrete} and \citet{paesleme2017gross} for surveys. We consider two applications: joint ventures with coordination costs and government formation with ministerial expertise. Although $M^\natural$-concavity need not survive arbitrary perturbations of the characteristic function, it is preserved by changes in the underlying parameters in these applications.

\subsection{Joint ventures with coordination costs}\label{sec-laminar}
Consider firms bargaining over membership and profit sharing in a joint venture requiring at least $q$ participants, with the profit subject to congestion and coordination costs. Let $w\in\mathbb R^n$, where $w_i$ is firm $i$'s exogenous contribution to the joint venture profit before accounting for congestion and coordination costs. Let $\mathcal L$ be a laminar family of subsets of $N$, meaning that any two members of $\mathcal L$ are disjoint or one contains the other. For each $A\in\mathcal L$, let $\phi_A:\{0,\dots,|A|\}\to\mathbb R$ have weakly decreasing increments. Given $q\in\{1,\dots,n\}$, let $\mathcal W=\{C\subseteq N:|C|\ge q\}$ and define
\[
    \chi(C)=\sum_{i\in C}w_i+\sum_{A\in\mathcal L}\phi_A(|C\cap A|),\qquad C\in\mathcal W,
\]
with parameters chosen so that $\chi(C)>0$ for all $C\in\mathcal W$. For example, taking $\phi_A=-g_A$, where $g_A(0)=0$ and $g_A$ has nonnegative, weakly increasing increments, gives congestion or coordination costs associated with the participating firms in group $A$. Firms then may be partitioned into local operating groups, which are in turn grouped into regions. The local groups, regions, and $N$ form a laminar family. The costs $g_A(|C\cap A|)$ can capture congestion at facilities shared by a local group or coordination among participating firms within a region, while $g_N(|C|)$ captures costs of coordinating the venture as a whole.

These characteristic functions are $M^\natural$-concave on $\mathcal W$ \citep[Section 3.6, item 4, and Theorem 3.2]{murota2016discrete}. Note that we allow $w_i<0$. Furthermore, a firm with $w_i>0$ can still lower coalition surplus when the additional coordination costs of including it exceed its contribution. Under Assumption ND, Theorem \ref{mnatural-unique} gives a unique equilibrium expected payoff vector. For fixed $q$ and $\mathcal L$, Corollary \ref{equilibrium-payoff-continuity} implies that this vector depends continuously on $w$, the values of the functions $\phi_A$, $p$, and $\delta$. Proposition \ref{max-value-no-delay} in Appendix \ref{appendix-no-delay-conditions} gives a sufficient condition for ND that allows non-monotone $\chi$.

With $\mathcal L=\{N\}$, coordination costs depend only on the number of participating firms. Taking $w_i=1$ for every $i\in N$ and $g_N(k)=\max\{k-1,0\}$ gives $\chi(C)=1$ for every feasible coalition, recovering the $q$-quota version of the model of \citet{baron1989bargaining}. With $\mathcal L=\varnothing$ and $w\in\mathbb R_{++}^n$, the characteristic function instead gives the additive $q$-quota environment of \citet{evdokimov2023equality}. Both cases have monotone $\chi$ and upward-closed $\mathcal W$, so Proposition \ref{monotone-no-delay} gives no delay and Theorem \ref{mnatural-unique} gives uniqueness. The general specification with $\mathcal L=\{N\}$ also includes the symmetric, nondecreasing, concave characteristic functions of Corollary~\ref{symmetric-concave-unique}.

\subsection{Government formation with ministerial expertise}\label{sec-assignment}
Consider a model of government formation in which a coalition's surplus depends on its members' expertise and their assignment to cabinet positions. Let the players be parties, let $T$ be a finite set of positions, and let $a_{it}\geq0$ be party $i$'s contribution to government surplus when it occupies position $t$. Assume that each party has a strictly positive contribution in at least one cabinet position. Given a quota $q\in\{1,\dots,n\}$, let $\mathcal W=\{C\subseteq N:|C|\ge q\}$ and define
\[
    \chi(C)=\max_M
    \sum_{(i,t)\in M}a_{it},
    \qquad C\in\mathcal W,
\]
where the maximum is taken over assignments $M\subseteq C\times T$ in which each party occupies at most one position and each position is assigned to at most one party. Coalition members without a position still contribute to meeting the quota, and unassigned positions contribute zero surplus. The assignment determines the total transferable surplus of the government, which is divided through bargaining.

This assignment formula defines an $M^\natural$-concave function on $\mathcal W$ \citep[Section 3.6, item 6, and Theorem 3.2]{murota2016discrete}.\footnote{Murota's formulation assigns every party in $C$ to a position. Adding $n$ dummy positions, each with zero surplus contribution for every party, makes that formulation equivalent to ours.} Our assumption on contributions ensures that $\chi(C)>0$ for every feasible coalition. Adding a party expands the available assignments, so the characteristic function is monotone. Proposition \ref{monotone-no-delay} gives no delay, and Theorem \ref{mnatural-unique} therefore yields a unique vector of stationary expected payoffs. Since each $\chi(C)$ is a maximum of finitely many linear functions of $(a_{it})$, Corollary \ref{equilibrium-payoff-continuity} implies that equilibrium expected payoffs depend continuously on $(a_{it})$, $p$, and $\delta$ for fixed $q$ and $T$.

This specification captures some important special cases. Let there be $n$ cabinet positions. Then a diagonal matrix $(a_{it})$ gives the additive $q$-quota environment again. In turn, with identical rows in the matrix $(a_{it})$, $\chi(C)$ equals the sum of the $|C|$ largest entries in the common row, so the characteristic function is symmetric, nondecreasing, and concave in coalition size \citep[Section~2]{paesleme2017gross}. This gives a special case of Corollary~\ref{symmetric-concave-unique}.

\section{Monotone comparative statics}\label{comparative-statics}

As noted in the introduction, uniqueness makes comparisons of equilibrium expected payoffs independent of equilibrium selection. Uniqueness alone, however, does not determine the direction of payoff changes. We show that the equitable grouping property also delivers comparative statics for a player's productivity and discount factor, without imposing further structural assumptions. The comparative statics allow the maximizing menus and supporting proposal distributions to change as the player's productivity or discount factor changes. The proofs are in Appendix \ref{appendix-comparative-statics}.

\subsection{Productivity}\label{comparative-statics-productivity}

The first comparative static fixes the collection of feasible coalitions $\mathcal W$ and perturbs the characteristic function $\chi$ as follows. Fix a player $i\in N$ and $a\in\mathbb R$. Define $\chi_a(C)=\chi(C)+a\mathbb I\{i\in C\}$ for all $C\in\mathcal W$, and let $\theta_a=(\chi_a,p,\delta)$.

If $\chi$ is $M^\natural$-concave, then so is $\chi_a$, because both transfers and replacements preserve the total number of occurrences of player $i$ in the two coalitions. In the joint venture application of Section \ref{sec-laminar}, this perturbation replaces $w_i$ with $w_i+a$ while holding coordination costs and productivities of other players fixed. In the government formation application of Section \ref{sec-assignment}, when there are $n$ cabinet positions, this perturbation replaces $a_{it}$ by $a_{it}+a$ for every position $t$, provided the shifted contributions remain nonnegative.

\begin{prop}\label{own-productivity}
    Fix $\theta=(\chi,p,\delta)\in\Theta$ with an $M^\natural$-concave characteristic function $\chi$, and fix $i\in N$. Let $I\subseteq\mathbb R$ be an interval such that $\theta_a\in\Theta^{\mathrm{ND}}$ for every $a\in I$, and let $y(a)$ be the unique equilibrium payoff vector under $\theta_a$. For any $a,a'\in I$ with $a<a'$, we have $0<y_i(a')-y_i(a)\leq a'-a$.
\end{prop}

Thus a player strictly benefits from a uniform increase in her productivity, and her payoff gain is at most the productivity increase.\footnote{\citet[Corollary~1]{gomes2015multilateral} derives local comparative statics with respect to coalition values and recognition probabilities using the inverse equilibrium Jacobian at regular, nondegenerate equilibria. He calculates the directions of these effects in specific quota and apex games. In his quota example, each player's payoff increases with her own contribution to coalition surplus. One can also compare players within one equilibrium. \citet{evdokimov2023equality} shows that, in the additive-quota model with uniform recognition probabilities and a common discount factor, more productive players have weakly higher expected payoffs.} If $\mathcal W$ is upward closed and $\chi$ is monotone, then $\chi_a$ is monotone for every $a\geq0$. Proposition \ref{monotone-no-delay} therefore ensures that the conclusion applies throughout $I=[0,\infty)$.

\subsection{Patience}\label{comparative-statics-patience}

The second comparative static fixes $\chi$, $p$, and the discount factors of all players other than $i$. For $d\in(0,1)$, write $\delta_i^d=d$, $\delta_j^d=\delta_j$ for $j\neq i$, and let $\theta^d=(\chi,p,\delta^d)$.

\begin{prop}\label{own-patience}
    Fix $\theta=(\chi,p,\delta)\in\Theta$ with an $M^\natural$-concave characteristic function $\chi$, and fix $i\in N$. Let $I\subseteq(0,1)$ be an interval such that $\theta^d\in\Theta^{\mathrm{ND}}$ for every $d\in I$, and let $y(d)$ be the unique equilibrium payoff vector under $\theta^d$. For any $d,d'\in I$ with $d<d'$, we have $y_i(d)\leq y_i(d')$, where equality holds if and only if $y(d)=y(d')$.
\end{prop}

A player's payoff is therefore weakly increasing in her own discount factor.\footnote{One can also compare players within one equilibrium. \citet[Corollary 2]{eraslan2016uniqueness} shows that, with uniform recognition probabilities in the Baron--Ferejohn model, $\delta_i\leq\delta_j$ implies $y_i\leq y_j$.} If her payoff is unchanged between two discount factors, every player's payoff is unchanged throughout the interval between them. If $\mathcal W$ is upward closed and $\chi$ is monotone, Proposition \ref{monotone-no-delay} ensures that the result applies throughout $I=(0,1)$.

The increase in expected payoff need not be strict, as illustrated by the following example.
\begin{expatiencerevisited}
    Keep the other primitives of Example~\ref{ex-two-player-ties} fixed and allow $\delta_1=d$ to vary. For every $d\in[\frac12,\frac23]$, the unique equilibrium payoff vector remains $y(d)=(1,1)$, at which player 1 proposes $\{1,2\}$, and player 2 proposes $\{1,2\}$ with probability $\frac2d-3$ and $\{2\}$ otherwise.
\end{expatiencerevisited}

Although Proposition~\ref{own-patience} rules out a decrease in $y_i(d)$, the player's ex ante expected payoff $\frac{y_i(d)}{d}$ can decrease. In the example above, player 1's ex ante expected payoff is $\frac1d$ and is strictly decreasing in $d\in[\frac12,\frac23]$. In the Baron--Ferejohn model with $2\leq q<n$, \citet[Theorem~1]{kawamori2005players} shows that a player's ex ante expected payoff is strictly decreasing in her own discount factor when recognition probabilities are sufficiently close to uniform and discount factors sufficiently close to a common value. His result also implies that, under these conditions, her expected payoff $y_i$ is strictly increasing in her own discount factor.\footnote{\citet{montero2015model} discusses this distinction in connection with a model of protocoalition bargaining in which increasing a player's probability of proposing within a formed coalition can reduce her ex ante expected payoff.}

\section{Conclusion}\label{conclusion}

In this paper, we show that proposer independence does not guarantee uniqueness of stationary expected payoffs in coalitional bargaining, even with strictly monotone characteristic functions that rule out equilibrium delay. Using fixed point index methods, we show how an equitable grouping property of maximizing menus yields uniqueness under no delay. We verify this property for $M^\natural$-concave characteristic functions and establish continuity of the unique equilibrium payoff vector. We also derive comparative statics for a player's productivity and discount factor. Applications to joint ventures and government formation illustrate these results.

In environments covered by Theorem~\ref{mnatural-unique}, uniqueness makes comparative statics independent of equilibrium selection. A natural next step is to study the effects of changes in the quota, recognition probabilities, and more general changes in coalition surplus. These questions may require additional structure, such as the threshold structure of recruitment used in the additive model of \citet{evdokimov2023equality}.

Another direction is to identify weaker restrictions on coalition surpluses and feasibility that ensure uniqueness under no delay. $M^\natural$-concavity is sufficient for the equitable grouping property to hold at all no-delay candidate payoff vectors. However, our uniqueness argument only uses the equitable grouping property in a neighborhood of every no-delay equilibrium payoff vector. Identifying the primitive conditions that guarantee this weaker requirement would therefore extend the result beyond $M^\natural$-concavity.

Turning to equilibrium delay, consider an equilibrium in which some players always pass and all remaining players have positive proposer surplus. Removing the passing players and all coalitions containing them yields a reduced game with adjusted recognition probabilities and discount factors. If the same partition holds across all equilibria and the common reduced game satisfies the hypotheses of Theorem~\ref{mnatural-unique}, equilibrium expected payoffs in the original game are unique. Identifying primitive conditions that ensure these requirements would extend our uniqueness result to some environments with delay. More difficult cases arise when the set of players who always pass differs across equilibria or when players with zero proposer surplus sometimes form non-empty coalitions. We leave these extensions for future work.

\begin{appendices}
\numberwithin{equation}{section}

\section{Sufficient conditions for no delay}\label{appendix-no-delay-conditions}

This appendix gives two further sufficient conditions for Assumption ND. The first concerns the standard monotone environment, in which feasible coalitions are upward closed and coalition surplus is monotone.

\begin{aprop}\label{monotone-no-delay}
    Fix $\theta=(\chi,p,\delta)\in\Theta$. Suppose that the feasible family $\mathcal W$ is upward closed, in the sense that $C\in\mathcal W$ and $C\subseteq D\subseteq N$ imply $D\in\mathcal W$, and that $\chi$ is monotone on $\mathcal W$, in the sense that $C,D\in\mathcal W$ and $C\subseteq D$ imply $\chi(C)\le \chi(D)$. Then $\theta\in\Theta^{\mathrm{ND}}$.
\end{aprop}

\begin{proof}
    Suppose that some player $i$ has $S_i(y;\theta)=0$ at an equilibrium payoff vector $y$. The equilibrium equation \eqref{payoff} then implies $y_i=0$. Moreover, not all players can have zero proposer surplus. Indeed, if $S_k(y;\theta)=0$ for every $k\in N$, then the equilibrium equations imply $y_k=0$ for every $k\in N$. But for any player $j$, the set $\mathcal W^j$ is non-empty by assumption, so choosing any coalition $C\in\mathcal W^j$ gives $S_j(y;\theta)\geq \chi(C)-y_C=\chi(C)>0$, a contradiction. Thus there exists some proposer $j$ with strictly positive proposer surplus. If $C\in\mathcal C_j(y;\theta)$, then upward-closed feasibility and monotonicity imply that player $i$ can propose $C\cup\{i\}$ and obtain proposer surplus at least $\chi(C)-y_C>0$, a contradiction.
\end{proof}

A second sufficient condition for Assumption ND compares the largest coalition surplus available to each player with the average of these maxima weighted by recognition probabilities. For a given primitive $\theta=(\chi,p,\delta)$, define $\chi_i^*=\max_{C\in\mathcal W^i}\chi(C)$ and $\overline\delta=\max_{i\in N}\delta_i$.

\begin{aprop}\label{max-value-no-delay}
    Fix primitives $\theta=(\chi,p,\delta)\in\Theta$. Suppose that $\chi_i^*>\overline\delta\sum_{j\in N}p_j\chi_j^*$ for every $i\in N$. Then $\theta\in\Theta^{\mathrm{ND}}$.
\end{aprop}

\begin{proof}
    Fix $y\in\mathcal E(\theta)$, and let $(\rho_j)_{j\in N}$ be equilibrium proposal distributions supporting $y$, extended to $\mathcal W^j\cup\{C_\varnothing\}$ by setting $\rho_j(C)=0$ for all $C\not\in\mathcal C_j(y)$. We first show that
    \begin{equation}\label{max-value-aggregate-bound}
        y_N\leq\overline\delta\sum_{j\in N}p_j\chi_j^*.
    \end{equation}
    By \eqref{problem}, $S_j(y)+y_C=\chi(C)$ for every non-empty $C\in\mathcal C_j(y)$, and $S_j(y)\rho_j(C_\varnothing)=0$. Thus, \eqref{payoff} gives
    \begin{equation}\label{max-value-payoff-decomposition}
        \begin{aligned}
            y_N\leq\overline\delta\sum_{i\in N}\frac{y_i}{\delta_i}&=\overline\delta\left[\sum_{i\in N}p_iS_i(y)+\sum_{i\in N}\sum_{j\in N}p_j\sum_{C\in\mathcal C_j^i(y)}\rho_j(C)y_i\right]\\
            &=\overline\delta\sum_{j\in N}p_j\left[S_j(y)+\sum_{C\in\mathcal W^j}\rho_j(C)y_C+\rho_j(C_\varnothing)y_N\right]\\
            &=\overline\delta\sum_{j\in N}p_j\left[\sum_{C\in\mathcal W^j}\rho_j(C)\chi(C)+\rho_j(C_\varnothing)y_N\right]\leq\overline\delta\sum_{j\in N}p_j\max\{\chi_j^*,y_N\}.
        \end{aligned}
    \end{equation}
    The first inequality uses $\delta_i\leq\overline\delta$ and $y_i\geq0$. The second inequality uses $\chi(C)\leq\chi_j^*$ for $C\in\mathcal W^j$ and the fact that the proposal probabilities sum to one.

    Suppose \eqref{max-value-aggregate-bound} does not hold. For every $j\in N$, we have $\chi_j^*>\overline\delta\sum_{k\in N}p_k\chi_k^*$ and therefore
    \begin{equation}\label{max-value-maximum-bound}
        \max\{\chi_j^*,y_N\}=\chi_j^*+y_N-\min\{\chi_j^*,y_N\}<\chi_j^*+y_N-\overline\delta\sum_{k\in N}p_k\chi_k^*,\qquad j\in N.
    \end{equation}
    Combining \eqref{max-value-payoff-decomposition} and \eqref{max-value-maximum-bound}, and using $\sum_{j\in N}p_j=1$, gives
    \[
        y_N\leq\overline\delta\sum_{j\in N}p_j\chi_j^*+\overline\delta\left(y_N-\overline\delta\sum_{j\in N}p_j\chi_j^*\right)<y_N,
    \]
    where the strict inequality follows from $\overline\delta<1$. This contradiction proves \eqref{max-value-aggregate-bound}.

    Finally, fix $i\in N$ and choose $C_i\in\mathcal W^i$ with $\chi(C_i)=\chi_i^*$. Since $y_{C_i}\leq y_N$, \eqref{max-value-aggregate-bound} and $\chi_i^*>\overline\delta\sum_{j\in N}p_j\chi_j^*$ imply
    \[
        S_i(y;\theta)\geq\chi_i^*-y_{C_i}\geq \chi_i^*-y_N\geq\chi_i^*-\overline\delta\sum_{j\in N}p_j\chi_j^*>0.
    \]
    Hence every equilibrium is no-delay.
\end{proof}

Proposition \ref{max-value-no-delay} can rule out delay even when adding a player reduces coalition surplus. Nonetheless, the test is conservative because it requires each player to belong to a feasible coalition whose surplus exceeds an upper bound on $y_N$, even though $\chi(C)>y_C$ can hold when $\chi(C)\leq y_N$.

Further, the condition becomes more demanding as $\overline\delta$ increases. For fixed $\chi$ and $p$, it is equivalent to
\[
    \overline\delta<\frac{\min_{i\in N}\chi_i^*}{\sum_{j\in N}p_j\chi_j^*}.
\]
Since every recognition probability is positive, the denominator strictly exceeds the numerator unless all $\chi_i^*$ are equal. Their largest value is $\chi^*=\max_{C\in\mathcal W}\chi(C)$. Thus, if some player belongs to no coalition attaining $\chi^*$, the condition fails when $\overline\delta$ is sufficiently close to one. If every player belongs to such a coalition, the condition holds for every $\overline\delta<1$.

Supplementary Appendix~\ref{appendix-limited-harm} gives an exact characterization of no delay through a \textit{limited harm} condition on reduced games. As shown in Supplementary Appendix~\ref{appendix-limited-harm-example}, in the setting of Observation~\ref{negative-externality-no-delay} this condition holds if and only if $a_1>\frac37$, compared with the bound $a_1>\frac67$ from Proposition~\ref{max-value-no-delay}.

\section{Proofs}\label{appendix}

\subsection{Proof of Proposition \ref{payoff-correspondence}}\label{appendix-payoff-correspondence}

\begin{proof}
    For $y\in Y$ and $\rho=(\rho_i)_{i\in N}$ with $\rho_i\in\Delta(\mathcal C_i(y))$, extend each $\rho_i$ to $\mathcal W^i\cup\{C_\varnothing\}$ by assigning probability zero to coalitions outside $\mathcal C_i(y)$, and define
    \[
        t_i(y,\rho)=\delta_i p_i\bigl(y_i+S_i(y)\bigr)+\delta_i\sum_{j\neq i}p_j\sum_{\substack{C\in\mathcal W^j\cup\{C_\varnothing\}:\\ C\ni i\text{ or }C=C_\varnothing}}\rho_j(C)y_i,\qquad i\in N.
    \]
    The definition of $R(y)$ in \eqref{aggregate inclusion-polytope-at-y} implies that $\mathcal T(y)$ defined in \eqref{correspondence} is exactly the set of all vectors $t(y,\rho)$ generated by such $\rho$.

    Suppose first that $\sigma$ is an equilibrium and let $y=v(\sigma)$. By \eqref{payoff}, there exist distributions $\rho_i$ supported on $\mathcal C_i(y)$ for each $i\in N$ such that $y=t(y,\rho)$. Hence $y\in\mathcal T(y)$.

    Conversely, suppose that $y\in\mathcal T(y)$. Choose distributions $\rho_i$ supported on $\mathcal C_i(y)$ such that $y=t(y,\rho)$. Construct a stationary strategy profile $\sigma$ as follows. Each proposer $i$ chooses a coalition $C_i\in\mathcal C_i(y)$ according to $\rho_i$. If $C_i=C_\varnothing$, she passes; otherwise she offers $y_j$ to each $j\in C_i\setminus\{i\}$, offers zero to players outside $C_i$, and keeps the remaining surplus. Each voter $i$ accepts a proposal if and only if it allocates at least $y_i$ to her. Let $d=\sum_{j\in N}p_j\rho_j(C_\varnothing)$ be the ex ante probability that $C_\varnothing$ is proposed under $\sigma$. Because all prescribed non-empty proposals are accepted, the payoffs induced by $\sigma$ satisfy
    \[
        v_i(\sigma) = t_i(y,\rho)+\delta_i d\bigl(v_i(\sigma)-y_i\bigr), \qquad i\in N.
    \]
    Since $t(y,\rho)=y$ and $\delta_i d<1$, it follows that $v(\sigma)=y$. By construction, every proposer $i$ randomizes only over coalitions in $\mathcal C_i(y)$, that is, over maximizers of $\chi(C)-y_C$, so no proposer can profitably deviate. The voting rule is optimal and weakly undominated because $y_i$ is the continuation payoff of player $i$. Hence $\sigma$ is an equilibrium.

    The correspondence $\mathcal T$ is non-empty valued. Moreover, $\mathcal T$ maps $Y$ into itself. To see this, first note that $t_i(y,\rho)\geq0$ for all $i\in N$. Further, if $C\in\mathcal C_i(y)$, then either $C=C_\varnothing$ and $y_i+S_i(y)=y_i$, or $C\neq C_\varnothing$ and $y_i+S_i(y)=y_i+\chi(C)-y_C=\chi(C)-\sum_{\ell\in C\setminus\{i\}}y_\ell\le \max_{D\in\mathcal W^i}\chi(D)$. Thus both $y_i+S_i(y)$ and $y_i$ belong to $Y_i$. Since the total probability weight attached to these terms is at most one, it follows that $t_i(y,\rho)\in Y_i$ for every $i\in N$.

    For the remaining properties of $\mathcal T$, note that $y\mapsto\Delta(\mathcal C_i(y))$ has compact convex values. Its values are exactly the distributions maximizing expected proposer surplus over the fixed simplex $\Delta(\mathcal W^i\cup\{C_\varnothing\})$, so $y\mapsto\Delta(\mathcal C_i(y))$ is upper hemicontinuous by the maximum theorem \cite[Theorem 17.31]{aliprantis2006infinite}. Since $t(y,\rho)$ is continuous in $(y,\rho)$ and affine in $\rho$, the induced correspondence $\mathcal T$ is compact valued, convex valued, and upper hemicontinuous by \citet[Theorem 17.23]{aliprantis2006infinite}. Kakutani's fixed point theorem \cite[Corollary 17.55]{aliprantis2006infinite} implies existence of $y\in Y$ such that $y\in\mathcal T(y)$.

    Finally, upper hemicontinuity and compact values imply that $\mathcal T$ has a closed graph. Therefore $\Fix(\mathcal T)$ is closed in the compact set $Y$, and hence is compact.
\end{proof}

\subsection{Proof of Lemma \ref{open}}\label{appendix-no-delay}

\begin{proof}
    \proofstep{The graph of the correspondence $\mathcal E:\Theta\rightrightarrows \mathbb R_+^n$ is closed.}\label{step-no-delay-closed-graph} Let $\theta^\ell\to\theta$ and $y^\ell\in \mathcal E(\theta^\ell)$ with $y^\ell\to y$. We want to show that $y\in \mathcal E(\theta)$. Since $y^\ell\in Y(\theta^\ell)$, passing to the limit in the inequalities defining $Y(\theta^\ell)$ gives $y\in Y(\theta)$.

    For each $\ell$, choose proposal distributions $\rho_i^\ell\in\Delta(\mathcal C_i(y^\ell;\theta^\ell))$, $i\in N$, supporting the payoff equations at $y^\ell$ under $\theta^\ell$. Extend each $\rho_i^\ell$ by assigning probability zero to coalitions outside $\mathcal C_i(y^\ell;\theta^\ell)$, so that $\rho_i^\ell\in\Delta(\mathcal W^i\cup\{C_\varnothing\})$. Since there are finitely many players, we can pass to a subsequence along which every $\rho_i^\ell$ converges. Denote the corresponding limits by $\rho_i$, $i\in N$.

    If $\rho_i(C)>0$, then $\rho_i^\ell(C)>0$ for all sufficiently large $\ell$, and therefore $\chi^\ell(C)-y_C^\ell\geq\chi^\ell(D)-y_D^\ell$ for every $D\in\mathcal W^i\cup\{C_\varnothing\}$. Passing to the limit, we obtain $C\in\mathcal C_i(y;\theta)$. Thus $\rho_i\in\Delta(\mathcal C_i(y;\theta))$ for every $i\in N$.

    Each surplus $S_i(y;\theta)$ is continuous in $(y,\theta)$, because it is the maximum of finitely many continuous functions. Taking limits in the payoff equations
    \[
        y_i^\ell=\delta_i^\ell p_i^\ell\bigl(y_i^\ell+S_i(y^\ell;\theta^\ell)\bigr)+\delta_i^\ell\sum_{j\neq i}p_j^\ell\sum_{\substack{C\in\mathcal W^j\cup\{C_\varnothing\}:\\ C\ni i\text{ or }C=C_\varnothing}}\rho_j^\ell(C)y_i^\ell,\qquad i\in N,
    \]
    implies that the limiting distributions satisfy the payoff equations at $y$ under $\theta$. Hence $y\in\mathcal T_\theta(y)$, and Proposition \ref{payoff-correspondence} gives $y\in\mathcal E(\theta)$.

    \proofstep{The set $\Theta^{\mathrm{ND}}$ is open in $\Theta$.} Take any sequence $\theta^\ell\in\Theta\setminus\Theta^{\mathrm{ND}}$ with $\theta^\ell\to\theta$. Fix a small neighborhood $\theta\in V\subset\Theta$ with compact closure. Define $\overline Y_i=[0,\sup_{\tilde\theta\in V}\max_{C\in\mathcal W^i}\tilde\chi(C)]$ for each $i\in N$, where $\tilde\theta=(\tilde\chi,\tilde p,\tilde\delta)$, and let $\overline Y=\overline Y_1\times\dots\times\overline Y_n$. Then $Y(\tilde\theta)\subseteq\overline Y$ for every $\tilde\theta\in V$, and $\overline Y$ is compact. After discarding finitely many terms, we can assume that $\theta^\ell\in V$ for every $\ell$. For each $\ell$, choose $y^\ell\in \mathcal E(\theta^\ell)$ and $i_\ell\in N$ such that $S_{i_\ell}(y^\ell;\theta^\ell)=0$. Since $N$ is finite and $y^\ell\in\overline Y$, after passing to a subsequence we can assume that $i_\ell=i$ for all $\ell$ and $y^\ell\to y\in\overline Y$. By Step \ref{step-no-delay-closed-graph}, $y\in \mathcal E(\theta)$, and by continuity, $S_i(y;\theta)=0$. Hence $\theta\notin\Theta^{\mathrm{ND}}$, so $\Theta\setminus\Theta^{\mathrm{ND}}$ is closed and its complement $\Theta^{\mathrm{ND}}$ is open.
\end{proof}

\subsection{Proof of Lemma \ref{mnatural-tie-structure}}\label{appendix-mnatural-tie-structure}

\begin{proof}
    If $y$ is a no-tie vector, then $L_{\mathcal A(y)}=\{0\}$ and the singleton groups $G_i=\{i\}$ satisfy both parts of the equitable grouping property. Therefore suppose that $y$ is a tie vector.  

    Let $f(C)=\chi(C)-y_C$ for $C\in\mathcal W$. Then $f$ is $M^\natural$-concave on $\mathcal W$. Moreover, because $y$ is no-delay, we have $\mathcal C_k(y)\subseteq\mathcal W^k$ for each $k\in K(y)$. For $C,D\in\mathcal W^k$, we have $k\in C\cap D$. Thus the transfers and replacements in Definition \ref{def-mnatural-concavity} preserve $k$'s membership in resulting coalitions, so the restriction of $f$ to $\mathcal W^k$ remains $M^\natural$-concave. By \citet[Theorem 3.6]{murota2016discrete}, each menu $\mathcal C_k(y)$ is therefore \textit{$M^\natural$-convex}: for any $C,D\in\mathcal C_k(y)$ and $i\in C\setminus D$, either $C-i,D+i\in\mathcal C_k(y)$, or there is $j\in D\setminus C$ such that $C-i+j,D+i-j\in\mathcal C_k(y)$.
  
    \proofstep{The group basis.}
    We first show that the span defining $L_{\mathcal A(y)}$ in
    \eqref{type-tie-difference-space} is unchanged if we only use single-player additions, removals, and replacements:
    \begin{equation}\label{eq:mnatural-elementary-span}
        L_{\mathcal A(y)}=\spa\{\mathbf1_C-\mathbf1_D\mid C,D\in\mathcal C_k(y),\ k\in K(y),\ |C\setminus D|\le1,\ |D\setminus C|\le1\}.
    \end{equation}
    The right-hand side of \eqref{eq:mnatural-elementary-span} is contained in $L_{\mathcal A(y)}$ by definition. For the reverse inclusion, fix $k\in K(y)$. We will show by induction on $m$ that $\mathbf1_C-\mathbf1_D$ belongs to the right-hand side of \eqref{eq:mnatural-elementary-span} for every $C,D\in\mathcal C_k(y)$ with $|C\triangle D|\leq m$, where $\triangle$ denotes symmetric difference. The case $m=0$ is immediate. Let $m\geq1$, suppose the claim holds for $m-1$, and take $C,D\in\mathcal C_k(y)$ with $|C\triangle D|=m$. Interchanging $C$ and $D$ if necessary, choose $i\in C\setminus D$. $M^\natural$-convexity of $\mathcal C_k(y)$ implies that either $E=C-i$ or $E=C-i+j$ belongs to $\mathcal C_k(y)$, with $j\in D\setminus C$ in the second case. Then we have $|E\triangle D|\leq m-1$ and $\mathbf1_C-\mathbf1_D=(\mathbf1_C-\mathbf1_E)+(\mathbf1_E-\mathbf1_D)$. Both differences on the right-hand side belong to the right-hand side of \eqref{eq:mnatural-elementary-span}, the first by construction and the second by induction, which proves the identity. 

    Next, note that the nonzero differences defining the span in \eqref{eq:mnatural-elementary-span} have the form $\pm e_i$ or $e_i-e_j$, where $e_i$ is the unit vector for player $i$. Discard every player $i$ such that $e_i\in L_{\mathcal A(y)}$. On the remaining players, define $i\sim j$ if $e_i-e_j\in L_{\mathcal A(y)}$. Since $L_{\mathcal A(y)}$ is a linear subspace, this is an equivalence relation. Let $G_1,\ldots,G_\ell$ be its equivalence classes.  By \eqref{eq:mnatural-elementary-span}, a vector $z$ belongs to $L_{\mathcal A(y)}^{\perp}$ if and only if it is zero on the discarded players and constant on each $G_t$. Hence $L_{\mathcal A(y)}^{\perp}=\spa\{\mathbf1_{G_1},\ldots,\mathbf1_{G_\ell}\}$. These indicators are linearly independent because the groups are non-empty and disjoint. This proves part (\ref{maximizing-group-basis}).

    \proofstep{Common recruitment counts.} For each $s$, the count $|C\cap G_s|$ is constant within every maximizing menu $\mathcal C_i(y)$ because $\mathbf1_{G_s}\in L_{\mathcal A(y)}^\perp$ and therefore $(\mathbf1_C-\mathbf1_D)^\top\mathbf1_{G_s}=|C\cap G_s|-|D\cap G_s|=0$ for all $C,D\in\mathcal C_i(y)$. It remains to show that these counts agree across proposers in the same group $G_t$ for each $t$.
    
    We first show that two players in $G_t$ connected by a replacement have equal proposer surpluses. Suppose that $C$ and $C-i+j$ belong to some menu $\mathcal C_k(y)$, with $i\in C$ and $j\notin C$. Necessarily $k\notin\{i,j\}$. Take any $D\in\mathcal C_i(y)$. If $j\in D$, then $S_j(y)\geq S_i(y)$. Otherwise, apply the exchange inequalities to coalitions $C-i+j$ and $D$ and player $j$. If \eqref{mnatural-transfer-inequality} applies, let $E=D+j$ and $F=C-i$. Otherwise, \eqref{mnatural-replacement-inequality} gives some $j'\in D\setminus(C-i+j)$, and we let $E=D-j'+j$ and $F=C-i+j'$. In either case, $F$ contains $k$, so $f(F)\leq S_k(y)$. The exchange inequality therefore gives $f(E)\geq f(D)=S_i(y)$, implying $S_j(y)\geq S_i(y)$. Reversing the roles of $i$ and $j$ gives $S_i(y)=S_j(y)$.

    We next show that recruitment counts agree between $\mathcal C_i(y)$ and $\mathcal C_j(y)$ for $i$ and $j$ chosen above. Take any $D\in\mathcal C_i(y)$. If $j\in D$, then $D$ also belongs to $\mathcal C_j(y)$ because $S_i(y)=S_j(y)$, so the counts agree because they are constant within each menu. Suppose instead that $j\notin D$, and construct $E$ and $F$ as above. The exchange inequality gives $f(E)+f(F)\geq f(D)+f(C-i+j)=S_i(y)+S_k(y)$. Since $f(E)\leq S_j(y)=S_i(y)$ and $f(F)\leq S_k(y)$, both bounds hold with equality. Thus $E\in\mathcal C_j(y)$ and $F\in\mathcal C_k(y)$. By construction, $\mathbf1_E-\mathbf1_D=\mathbf1_{C-i+j}-\mathbf1_F\in L_{\mathcal A(y)}$, so $|E\cap G_s|=|D\cap G_s|$ for every $s$. Because recruitment counts are constant within each individual maximizing menu, it follows that the recruitment counts are also constant between $\mathcal C_i(y)$ and $\mathcal C_j(y)$.

    Identity \eqref{eq:mnatural-elementary-span} implies any two distinct players in $G_t$ are connected by a chain of replacements, each involving two coalitions in the same maximizing menu. To see this, suppose otherwise and choose a non-empty proper subset $H\subset G_t$ with no replacement connecting $H$ to $G_t\setminus H$. Since $\mathbf1_{G_t}\in L_{\mathcal A(y)}^\perp$, no addition or removal within a maximizing menu involves a player in $G_t$, and no replacement connects $G_t$ to its complement. Thus $\mathbf1_H$ is orthogonal to every difference on the right-hand side of \eqref{eq:mnatural-elementary-span}, so $\mathbf1_H\in L_{\mathcal A(y)}^\perp$. This contradicts Step 1, because $\mathbf1_H$ is not constant on $G_t$. Repeating the above argument along such a chain of replacements, we conclude that for every $t,s$ the count $|C\cap G_s|$ is the same for all $i\in G_t$ and $C\in\mathcal C_i(y)$, proving part (ii).
\end{proof}

\subsection{Proof of Lemma \ref{structural-local-index}}\label{appendix-structural-local-index}

\begin{proof}
Fix $\theta\in\Theta$ and suppose that $\chi$ is $M^\natural$-concave on $\mathcal W$. Further, fix $y_0\in\mathcal E(\theta)$ and suppose that $y_0$ is a no-delay vector of expected payoffs. If $y_0$ is no-tie, the conclusion follows from Lemma \ref{notie-local-degree}. Henceforth in this subsection, assume that $K(y_0)\neq\varnothing$.\footnote{As noted in Section \ref{appendix-comparative-statics}, the argument below also applies when $K(y_0)=\varnothing$. Lemma \ref{notie-local-degree} applies without $M^\natural$-concavity or no delay and, together with Lemma \ref{local-degree-criterion}, shows that payoff multiplicity requires at least one equilibrium payoff vector with ties, as explained at the end of Section \ref{deterministic-jacobian}. Its proof also introduces the ordering and determinant argument adapted in the proof of Lemma \ref{structural-local-index}.}

\proofstep{Local restrictions and homotopy.}\label{step-structural-local-setup}
By Proposition \ref{payoff-correspondence}, we have $y_0\in\mathcal T(y_0)$. Hence \eqref{correspondence} implies that there exists $r_0\in R(y_0)$ such that
\begin{equation}\label{structural-local-base-payoffs}
    y_{0,i}=\delta_i p_iS_i(y_0)+\delta_i\bigl(p_{C^i(y_0)}+r^i_0\bigr)y_{0,i},\qquad i\in N.
\end{equation}
Since $p_{C^i(y_0)}+r_0^i\leq1$, $S_i(y_0)>0$ implies $y_{0,i}>0$. Solving \eqref{structural-local-base-payoffs} then gives
\begin{equation}\label{structural-local-base-aggregate}
    r_0^i=\frac1{\delta_i}-p_{C^i(y_0)}-\frac{p_iS_i(y_0)}{y_{0,i}},\qquad i\in N,
\end{equation}
which determines $r_0$ uniquely. We call $r_0$ the aggregate inclusion vector \textit{supporting} $y_0$. Moreover, $S_i(y_0)>0$ implies that every $C\in\mathcal C_i(y_0)$ is non-empty, so  $y_{0,i}\leq y_{0,C}<\chi(C)\leq\max_{D\in\mathcal W^i}\chi(D)$ for every $i\in N$. It follows that $y_0\in\operatorname{int}Y$.

Strictly suboptimal coalitions at $y_0$ remain strictly suboptimal in its neighborhood. We can therefore choose an open neighborhood $V$ of $y_0$ with $\overline V\subset\operatorname{int}Y$ such that $S_i(y)>0$ and $\mathcal C_i(y)\subseteq\mathcal C_i(y_0)$ for every $i\in N$ and $y\in V$. Specifically, each proposer outside $K(y_0)$ still has the same unique maximizing coalition, and each proposer in $K(y_0)$ can have some (or all) original ties broken. Write $K_0=K(y_0)$ and keep this set of originally tied proposers fixed for the remainder of this proof.

Since all coalitions in $\mathcal C_i(y_0)$ are tied at $y_0$, for $y\in V$ we have
\[
    \chi(C)-y_C=S_i(y_0)-(y-y_0)_C,
    \qquad C\in\mathcal C_i(y_0).
\]
Therefore, within $\mathcal C_i(y_0)$, maximizing $\chi(C)-y_C$ is equivalent to minimizing $(y-y_0)_C=(y-y_0)^\top\mathbf1_C$, and $\mathcal C_i(y)$ consists exactly of the coalitions attaining this minimum. Define
\begin{equation}\label{structural-local-aggregate-set}
    F(y)=\left\{r\in[0,1]^n\mid r=\sum_{k\in K_0}p_k\sum_{C\in\mathcal C_k(y)}\rho_k(C)\mathbf1_C,\ \rho_k\in\Delta(\mathcal C_k(y)),\ k\in K_0\right\}\subseteq R(y_0),
\end{equation}
and note that for any $r\in F(y)$, player $i$'s total inclusion probability is $p_{C^i(y_0)}+r^i$.\footnote{Note that $F(y)$ generally differs from the aggregate inclusion polytope $R(y)$ because the former uses $K_0=K(y_0)$ while the latter uses $K(y)$.}

The convex-hull representation in \eqref{aggregate-inclusion-convex-hull} expresses $R(y_0)$ as a weighted sum of convex hulls of coalition incidence vectors. In this representation, restricting each menu $\mathcal C_k(y_0)$ to $\mathcal C_k(y)$ gives the set $F(y)$. We will show that this restriction selects exactly the minimizing aggregates:
\begin{equation}\label{structural-local-minimizing-face}
    F(y)=\argmin_{r\in R(y_0)}(y-y_0)^\top r,
    \qquad y\in V.
\end{equation}

First, take $r\in F(y)$ and proposal distributions $\rho_k\in\Delta(\mathcal C_k(y))$ generating it. If $r$ did not minimize $(y-y_0)^\top r$ over $R(y_0)$, then for some $k\in K_0$, a coalition $C$ with $\rho_k(C)>0$ would fail to minimize $(y-y_0)^\top\mathbf1_C$ over $\mathcal C_k(y_0)$. This contradicts $C\in\mathcal C_k(y)$. Conversely, any aggregate minimizing $(y-y_0)^\top r$ over $R(y_0)$ can be generated by proposal distributions over the original menus $\mathcal C_k(y_0)$. None of these distributions can assign positive probability to a nonminimizing coalition because $p_k>0$ implies that shifting this probability to a minimizing coalition would strictly lower the objective. Thus these distributions are supported on $\mathcal C_k(y)$. In particular, $F(y)$ is a non-empty face of $R(y_0)$ for every $y\in V$.

Now, consider the homotopy
\begin{equation}\label{structural-local-surplus-homotopy}
    H_\lambda(y)=(1-\lambda)y_0+\lambda\mathcal T(y),\qquad y\in V,\quad \lambda\in[0,1].
\end{equation}
Then, $H_0$ is the constant map $y_0$, while $H_1=\mathcal T$. Since $y_0\in\mathcal T(y_0)$, the vector $y_0$ is a fixed point of every $H_\lambda$. We will show that it is the only fixed point throughout this homotopy in a common neighborhood.

\proofstep{Restrictions at nearby fixed points of the homotopy.}\label{step-structural-local-faces}
At $\lambda=0$, the only fixed point is $y_0$. Fix $\lambda\in(0,1]$ and $y\in V$ with $y\in H_\lambda(y)$, and let $r\in F(y)$ generate the corresponding element of $\mathcal T(y)$. Subtracting $\lambda$ times \eqref{structural-local-base-payoffs} from the homotopy fixed point equation $y_i=(1-\lambda)y_{0,i}+\lambda\bigl[\delta_ip_iS_i(y)+\delta_i\bigl(p_{C^i(y_0)}+r^i\bigr)y_i\bigr]$ and rearranging gives
\begin{equation}\label{structural-local-aggregate-change}
    \lambda(r^i-r_0^i)=\frac{1/\delta_i-\lambda[p_{C^i(y_0)}+r_0^i]}{y_i}(y_i-y_{0,i})+\lambda\frac{p_i}{y_i}\bigl[S_i(y_0)-S_i(y)\bigr],\qquad i\in N.
\end{equation}
By \eqref{structural-local-minimizing-face}, $r\in F(y)$ and $r_0\in R(y_0)$ imply $(y-y_0)^\top(r-r_0)\leq0$. Solving \eqref{structural-local-aggregate-change} for $(y_i-y_{0,i})/\lambda$, multiplying by $r^i-r_0^i$, and summing over players therefore gives
\begin{equation}\label{structural-local-fixed-point-condition}
    0\geq\frac{(y-y_0)^\top(r-r_0)}{\lambda}
    =\sum_{i\in N}\frac{y_i(r^i-r_0^i)^2+p_i[S_i(y)-S_i(y_0)](r^i-r_0^i)}{1/\delta_i-\lambda[p_{C^i(y_0)}+r_0^i]}.
\end{equation}
The denominators satisfy $1/\delta_i-\lambda[p_{C^i(y_0)}+r_0^i]\geq1/\delta_i-1>0$ because $p_{C^i(y_0)}+r_0^i\leq1$ and $\delta_i<1$.

We can now further shrink $V$ so that $r_0\in F(y)$ at every fixed point of every $H_\lambda$ in $V$. Otherwise, there would be sequences $\lambda^\ell\in(0,1]$ and $y^\ell\in H_{\lambda^\ell}(y^\ell)$ with $y^\ell\to y_0$ and $r_0\notin F(y^\ell)$, supported by aggregates $r^\ell\in F(y^\ell)$. Since $R(y_0)$ is compact and has finitely many faces, we can pass to a subsequence on which $F(y^\ell)$ is constant, $r^\ell\to r$ in this closed face, and $\lambda^\ell\to \lambda\in[0,1]$. Passing to the limit in \eqref{structural-local-fixed-point-condition} and using continuity of the proposer surpluses gives
\[
    0\geq\sum_{i\in N}\frac{y_{0,i}(r^i-r_0^i)^2}{1/\delta_i-\lambda[p_{C^i(y_0)}+r_0^i]}.
\]
Every coefficient is strictly positive, so $r=r_0$. Thus $r_0\in F(y^\ell)$, a contradiction.

For any such fixed point $y$ supported by $r$, we now have $r,r_0\in F(y)$. Choose proposal distributions $\rho_k,\widetilde\rho_k\in\Delta(\mathcal C_k(y))$, $k\in K_0$, generating $r$ and $r_0$, respectively, as in \eqref{structural-local-aggregate-set}. For $k\in K_0\setminus K(y)$, both distributions assign probability one to $C_k(y)$, so proposer $k$ contributes $p_k\mathbf1_{C_k(y)}$ to both vectors. These contributions cancel in $r-r_0$. For each $k\in K(y)$, fix any $D\in\mathcal C_k(y)$. Since both distributions sum to one, proposer $k$'s contribution to $r-r_0$ satisfies
\[
    p_k\sum_{C\in\mathcal C_k(y)}[\rho_k(C)-\widetilde\rho_k(C)]\mathbf1_C=p_k\sum_{C\in\mathcal C_k(y)}[\rho_k(C)-\widetilde\rho_k(C)](\mathbf1_C-\mathbf1_D).
\]
By \eqref{type-tie-difference-space}, every difference $\mathbf1_C-\mathbf1_D$ on the right-hand side belongs to $L_{\mathcal A(y)}$. Summing over proposers gives $r-r_0\in L_{\mathcal A(y)}$.

Moreover, every tie in $\mathcal A(y)$ holds at both $y$ and $y_0$, so $(y-y_0)_C=(y-y_0)_D$ whenever $C,D\in\mathcal C_k(y)$. By \eqref{type-tie-difference-space}, $y-y_0\in L_{\mathcal A(y)}^{\perp}$. Thus, for every $y\in V$ and $\lambda\in(0,1]$ with $y\in H_\lambda(y)$, we have
\begin{equation}\label{structural-local-face-directions}
    r-r_0\in L_{\mathcal A(y)}, \qquad y-y_0\in L_{\mathcal A(y)}^{\perp}.
\end{equation}

\proofstep{Group equations.}\label{step-structural-local-groups}
We want to show that $y_0$ is the only fixed point of every $H_\lambda$ in $V$ chosen above. Fix $\lambda\in(0,1]$ and a fixed point $y\in V$, and let $r\in F(y)$ generate $y$ as in Step \ref{step-structural-local-faces}. Since $\chi$ is $M^\natural$-concave and every proposer surplus at $y$ is positive, Lemma \ref{mnatural-tie-structure} gives an equitable grouping $G_1,\ldots,G_\ell$ at $y$. Let $X=[\mathbf1_{G_1}\ \cdots\ \mathbf1_{G_\ell}]\in\mathbb R^{n\times\ell}$. By part (\ref{maximizing-group-basis}), the columns of $X$ form a basis of $L_{\mathcal A(y)}^{\perp}$. Thus \eqref{structural-local-face-directions} gives $X^\top(r-r_0)=0$ and $y-y_0=X\xi$ for some $\xi\in\mathbb R^\ell$.

If $\ell=0$, then $y=y_0$ and we are done. Suppose that $\ell>0$. Because the groups are disjoint, $y_j-y_{0,j}=\xi_s$ for $j\in G_s$, and $y_j=y_{0,j}$ outside these groups. For $i\in G_t$, choose $D_i\in\mathcal C_i(y)$ and write $c_{ts}=|D_i\cap G_s|$. By part (\ref{maximizing-recruitment-counts}), $|D_i\cap G_s|$ is independent of both the proposer $i\in G_t$ and the choice of $D_i$ and therefore $c_{ts}$ is well-defined. Define the count matrix $\mathsf C=(c_{ts})_{t,s=1}^{\ell}$.

Since $D_i\in\mathcal C_i(y)\subseteq\mathcal C_i(y_0)$, we have $S_i(y_0)=\chi(D_i)-y_{0,D_i}$ and $S_i(y)=\chi(D_i)-y_{D_i}$. Using the group coordinates and constant counts, for $i\in G_t$ we get 
\[
    S_i(y_0)-S_i(y)=(y-y_0)_{D_i}=\sum_{s=1}^{\ell}|D_i\cap G_s|\xi_s=\sum_{s=1}^{\ell}c_{ts}\xi_s=(\mathsf C\xi)_t.
\]
In turn, since $X^\top(r-r_0)=0$, we have $\mathbf1_{G_t}^\top(r-r_0)=\sum_{i\in G_t}(r^i-r_0^i)=0$. Therefore, summing \eqref{structural-local-aggregate-change} over $i\in G_t$, we get
\[
    0=\lambda\sum_{i\in G_t}(r^i-r_0^i)=A_t\xi_t+\lambda B_t(\mathsf C\xi)_t,
\]
where
\[
    A_t=\sum_{i\in G_t}\frac{1/\delta_i-\lambda[p_{C^i(y_0)}+r_0^i]}{y_i}>0, \qquad B_t=\sum_{i\in G_t}\frac{p_i}{y_i}>0.
\]
Because these expressions hold for all $t=1,\dots,\ell$, we can write them in matrix form
\begin{equation}\label{structural-local-group-equations}
    M\xi=0, \qquad M=\operatorname{diag}(A_t)_{t=1}^{\ell} +\lambda\operatorname{diag}(B_t)_{t=1}^{\ell}\mathsf C.
\end{equation}

\proofstep{Isolation.}\label{step-structural-local-isolation}
We will now show that $M$ is nonsingular by adapting the ordering and determinant argument from Steps \ref{step-notie-order}--\ref{step-notie-determinant} of the proof of Lemma \ref{deterministic-positive-det} to the groups $G_1,\dots,G_\ell$. For each group $t$, let $\overline S_t=\max_{i\in G_t}S_i(y)$. We will show that $c_{ts}>0$ implies $\overline S_s\geq\overline S_t$, with equality only when the rows $t$ and $s$ of the count matrix $\mathsf C$ coincide. If $c_{ts}>0$, choose $i\in G_t$ attaining $\overline S_t$ and a player $j\in D_i\cap G_s$. The coalition $D_i$ is also feasible for proposer $j$, so proposer independence gives
\[
    \overline S_s\geq S_j(y)\geq\chi(D_i)-y_{D_i}=S_i(y)=\overline S_t.
\]
If $\overline S_s=\overline S_t$, all inequalities are equalities, and $D_i$ also belongs to $\mathcal C_j(y)$. Part (\ref{maximizing-recruitment-counts}) then gives $c_{tv}=|D_i\cap G_v|=c_{sv}$ for every $v=1,\ldots,\ell$.

Partition the group indices into classes having the same value of $\overline S_t$ and the same row of $\mathsf C$. Order these classes by increasing $\overline S_t$, with arbitrary order between distinct classes having the same value. If the class of $s$ precedes that of $t$, then $\overline S_s\leq\overline S_t$. If $c_{ts}>0$, the preceding argument gives equality and identical count rows, contradicting that these classes are distinct. Hence $c_{ts}=0$. Thus $\mathsf C$ is block upper triangular, with identical rows within each diagonal block. The same ordering makes $M$ block upper triangular.

For each diagonal block $Q$ and $s\in Q$, let $c_{Qs}$ denote the common value of $c_{ts}$ over $t\in Q$, and collect these counts in the column vector $c_Q=(c_{Qs})_{s\in Q}$. Then $\mathsf C_Q=\mathbf1_{|Q|}c_Q^\top$ and $M_Q=\operatorname{diag}(A_t)_{t\in Q}+\lambda(B_t)_{t\in Q}c_Q^\top$. The matrix determinant lemma gives $\det M_Q=\left(\prod_{t\in Q}A_t\right)\left(1+\lambda\sum_{t\in Q}c_{Qt}\frac{B_t}{A_t}\right)>0$. Thus $\det M>0$, so \eqref{structural-local-group-equations} implies $\xi=0$ and hence $y=y_0$. The same conclusion holds at $\lambda=0$, when $H_0$ is the constant map $y_0$. Thus every $H_\lambda$, $\lambda\in[0,1]$, has exactly this fixed point in $V$. In particular, $y_0$ is isolated.

\proofstep{Local index.}\label{step-structural-local-index}
\interlinepenalty=10000
Let $U$ be any bounded open neighborhood of $y_0$ with $\overline U\subset V$. By Proposition \ref{payoff-correspondence} and convexity of $Y$, the correspondence $(\lambda,y)\mapsto H_\lambda(y)$ is jointly upper hemicontinuous with non-empty compact convex values in $Y$. Each $H_\lambda|_{\overline U}$ has no fixed point on the boundary of $U$. Homotopy invariance and Normalization therefore give $\Lambda_Y(\mathcal T|_{\overline U})=\Lambda_Y(H_0|_{\overline U})=+1$.
\end{proof}

\subsection{Proofs of Propositions \ref{own-productivity} and \ref{own-patience}}\label{appendix-comparative-statics}

The local construction and group calculations in the proof of Lemma \ref{structural-local-index} continue to hold when $y_0$ is a no-delay equilibrium payoff vector without ties. In this case, $K(y_0)=\varnothing$ and $r_0=0$, and every sufficiently nearby candidate payoff vector $y$ has the same unique maximizing coalitions as $y_0$. Hence $F(y)=\{0\}$ and $L_{\mathcal A(y)}=\{0\}$ throughout that neighborhood. The two inclusions in \eqref{structural-local-face-directions} therefore hold automatically with $r=r_0=0$. The group representation in Step \ref{step-structural-local-groups} then applies with the singleton groups $G_i=\{i\}$, $i\in N$.

The next lemma extends the argument in Step \ref{step-structural-local-isolation} of the proof of Lemma \ref{structural-local-index} by giving bounds on the diagonal entries of the inverse of a matrix with the same structure.

\begin{alem}\label{comparative-group-matrix}
    Fix $\theta\in\Theta$ and $y\in Y(\theta)$ with $S_j(y;\theta)>0$ for every $j\in N$. Suppose that $y$ satisfies the equitable grouping property with groups $G_1,\ldots,G_\ell$, where $\ell\geq1$. For $j\in G_t$ and $C\in\mathcal C_j(y;\theta)$, let $c_{ts}=|C\cap G_s|$ and write $\mathsf C=(c_{ts})_{t,s=1}^{\ell}$. For any $A_t,B_t>0$, define $M=\operatorname{diag}(A_t)_{t=1}^{\ell}+\operatorname{diag}(B_t)_{t=1}^{\ell}\mathsf C$. Then $M$ is nonsingular and
    \begin{equation}\label{comparative-inverse-diagonal-bound}
        0<(M^{-1})_{gg}<\frac1{A_g}, \qquad g=1,\ldots,\ell.
    \end{equation}
\end{alem}

\begin{proof}
    The counts $c_{ts}$ are well-defined by part (\ref{maximizing-recruitment-counts}) of Definition \ref{def-group-structure}. By the ordering argument in Step \ref{step-structural-local-isolation} of the proof of Lemma \ref{structural-local-index}, $\mathsf C$ and $M$ are block upper triangular, with identical rows within each diagonal block of $\mathsf C$. For each diagonal block $Q$ and $s\in Q$, let $c_{Qs}$ denote the common value of $c_{ts}$ over $t\in Q$, and collect these counts in the column vector $c_Q=(c_{Qs})_{s\in Q}$. Then $M_Q=\operatorname{diag}(A_t)_{t\in Q}+(B_t)_{t\in Q}c_Q^\top$. By the matrix determinant lemma, we have
    \begin{equation}\label{comparative-block-determinant}
        \det M_Q=\left(\prod_{t\in Q}A_t\right)\left(1+\sum_{t\in Q}c_{Qt}\frac{B_t}{A_t}\right)>0.
    \end{equation}
    Thus every diagonal block is nonsingular, and therefore $M$ is nonsingular.

    Let $Q$ be the diagonal block containing $g$. Since $M$ is block upper triangular with nonsingular diagonal blocks, $M^{-1}$ is block upper triangular with $(M^{-1})_{QQ}=M_Q^{-1}$. By the adjugate formula, $(M_Q^{-1})_{gg}$ equals the determinant of the matrix obtained by deleting row and column $g$ from $M_Q$, divided by $\det M_Q$. Using \eqref{comparative-block-determinant} for the denominator and the analogous formula for the numerator gives
    \[
        (M^{-1})_{gg}=(M_Q^{-1})_{gg}=\frac1{A_g}\frac{1+\sum_{t\in Q\setminus\{g\}}c_{Qt}B_t/A_t}{1+\sum_{t\in Q}c_{Qt}B_t/A_t}.
    \]
    Every maximizing coalition of a proposer in $G_g$ contains that proposer, so $c_{Qg}=c_{gg}\geq1$, which implies \eqref{comparative-inverse-diagonal-bound}.
\end{proof}

\subsubsection{Proof of Proposition \ref{own-productivity}}\label{appendix-own-productivity}

\begin{proof}
    Fix $a_0\in I$, and write $\theta_0=\theta_{a_0}$ and $y_0=y(a_0)$. We suppress $\theta_0$ in the notation for maximizing menus, proposer surpluses, types, and inclusion polytopes. Put $K_0=K(y_0)$, and let $r_0\in R(y_0)$ be the unique supporting aggregate inclusion vector given by \eqref{structural-local-base-aggregate}. As shown at the start of Section \ref{appendix-structural-local-index}, $y_0\in\operatorname{int}Y(\theta_0)$.

    \proofstep{Restrictions at nearby equilibria.}\label{step-productivity-menus}
    Following the construction in Step \ref{step-structural-local-setup} of the proof of Lemma \ref{structural-local-index}, choose a neighborhood $V\subset\operatorname{int}Y(\theta_0)$ of $y_0$ such that $S_j(z)>0$ and $\mathcal C_j(z)\subseteq\mathcal C_j(y_0)$ for every $j\in N$ and $z\in V$.

    For $a_0+\varepsilon\in I$, put $y=y(a_0+\varepsilon)$. To express the maximizing menus and proposer surpluses at $y$ using the baseline characteristic function $\chi_{a_0}$, define the auxiliary candidate payoff vector $z=y-\varepsilon e_i$, where $e_i$ is the $i$th standard basis vector in $\mathbb R^n$. For every $C\in\mathcal W$,
    \begin{equation}\label{productivity-menu-translation}
        \chi_{a_0+\varepsilon}(C)-y_C=\chi_{a_0}(C)-z_C.
    \end{equation}
    Thus the maximizing menus and proposer surpluses at $y$ under $\theta_{a_0+\varepsilon}$ are exactly those at $z$ under $\theta_0$. We will obtain the desired payoff comparison by bounding $z_i-y_{0,i}$ and using $y_i-y_{0,i}=\varepsilon+(z_i-y_{0,i})$. Corollary~\ref{equilibrium-payoff-continuity} gives $y\to y_0$ and $z\to y_0$ as $\varepsilon\to0$, so $z\in V$ for all sufficiently small $\varepsilon$.

    Define $F(z)$ using \eqref{structural-local-aggregate-set}, with $z$ in place of $y$, so that $F(z)$ is a non-empty face of $R(y_0)$ by \eqref{structural-local-minimizing-face}. Let $r$ be the vector generated by the equilibrium proposal distributions of the proposers in $K_0$ at $y$ under $\theta_{a_0+\varepsilon}$. By \eqref{productivity-menu-translation}, $r\in F(z)$, and proposers outside $K_0$ retain their unique maximizing coalitions. The equilibrium payoff equations are
    \[
        y_j=\delta_jp_jS_j(z)+\delta_j\bigl[p_{C^j(y_0)}+r^j\bigr]y_j, \qquad j\in N.
    \]
    Solving for $r^j$ and subtracting \eqref{structural-local-base-aggregate} gives
    \begin{equation}\label{productivity-aggregate-change}
        r^j-r_0^j=\frac{p_jS_j(y_0)}{y_{0,j}y_j}(y_j-y_{0,j}) +\frac{p_j}{y_j}\bigl[S_j(y_0)-S_j(z)\bigr], \qquad j\in N.
    \end{equation}
    Since $y,z\to y_0$ as $\varepsilon\to0$, this identity implies $r\to r_0$. The remaining arguments in Step \ref{step-structural-local-faces} of the proof of Lemma \ref{structural-local-index} therefore give, for all sufficiently small $\varepsilon$,
    \begin{equation}\label{comparative-face-directions}
        r-r_0\in L_{\mathcal A(z)},\qquad z-y_0\in L_{\mathcal A(z)}^\perp.
    \end{equation}

    \proofstep{Local payoff bounds.}\label{step-productivity-groups}
    Since $\chi_{a_0}$ is $M^\natural$-concave and $S_j(z)>0$ for every $j\in N$, Lemma \ref{mnatural-tie-structure} gives an equitable grouping $G_1,\ldots,G_\ell$ at $z$. Let $X=[\mathbf1_{G_1}\ \cdots\ \mathbf1_{G_\ell}]$. By part (\ref{maximizing-group-basis}) and \eqref{comparative-face-directions}, we have $X^\top(r-r_0)=0$ and
    \[
        z-y_0=X\xi,\qquad y-y_0=X\xi+\varepsilon e_i
    \]
    for some $\xi\in\mathbb R^\ell$. Note that the groups, and therefore $X$ and $\xi$, can depend on $\varepsilon$. If $\ell=0$, then $z=y_0$ and $y_i-y_{0,i}=\varepsilon$. Suppose that $\ell>0$, and let $\mathsf C$ be the count matrix at $z$ defined in Lemma \ref{comparative-group-matrix}. The surplus calculation in Step \ref{step-structural-local-groups} of the proof of Lemma \ref{structural-local-index} gives $S_j(y_0)-S_j(z)=(\mathsf C\xi)_t$ for $j\in G_t$.
    Because the groups are disjoint, $y_j-y_{0,j}=\xi_t+\varepsilon\mathbb I\{j=i\}$ for $j\in G_t$. Summing \eqref{productivity-aggregate-change} over $j\in G_t$ and using $X^\top(r-r_0)=0$ therefore gives
    \begin{equation}\label{productivity-group-equations}
        0=\sum_{j\in G_t}(r^j-r_0^j)=A_t\xi_t+B_t(\mathsf C\xi)_t+\varepsilon\alpha_i\mathbb I\{i\in G_t\}, \qquad t=1,\ldots,\ell,
    \end{equation}
    where
    \[
        A_t=\sum_{j\in G_t}\frac{p_jS_j(y_0)}{y_{0,j}y_j}>0, \qquad B_t=\sum_{j\in G_t}\frac{p_j}{y_j}>0, \qquad \alpha_i=\frac{p_iS_i(y_0)}{y_{0,i}y_i}>0.
    \]
    Let $M=\operatorname{diag}(A_t)_{t=1}^{\ell}+\operatorname{diag}(B_t)_{t=1}^{\ell}\mathsf C$. If $i$ does not belong to any group, \eqref{productivity-group-equations} becomes $M\xi=0$. Lemma \ref{comparative-group-matrix} gives $\xi=0$, so $z=y_0$ and $y_i-y_{0,i}=\varepsilon$.

    If $i\in G_g$, the same equations give $M\xi=-\varepsilon\alpha_i e_g$, where $e_g$ is the $g$th standard basis vector in $\mathbb R^\ell$. Hence
    \[
        y_i-y_{0,i}=\varepsilon+\xi_g=\varepsilon\bigl[1-\alpha_i(M^{-1})_{gg}\bigr].
    \]
    Since $0<\alpha_i\leq A_g$, the bound \eqref{comparative-inverse-diagonal-bound} implies $0<1-\alpha_i(M^{-1})_{gg}<1$. Combining the cases, including $\ell=0$, gives
    \begin{equation}\label{productivity-local-comparison}
        0<\frac{y_i(a_0+\varepsilon)-y_i(a_0)}{\varepsilon}\leq1
    \end{equation}
    for every sufficiently small nonzero $\varepsilon$ with $a_0+\varepsilon\in I$.

    \proofstep{Comparison on the interval.}
    For small positive $\varepsilon$, \eqref{productivity-local-comparison} gives $y_i(a_0+\varepsilon)>y_i(a_0)$ and $(a_0+\varepsilon)-y_i(a_0+\varepsilon)\geq a_0-y_i(a_0)$. Because $a_0\in I$ was arbitrary, these local comparisons and continuity imply that $a\mapsto y_i(a)$ is strictly increasing and $a\mapsto a-y_i(a)$ is nondecreasing on $I$.\footnote{\interlinepenalty=10000\label{fn-local-to-global-monotonicity}Let $f$ be continuous on an interval $I$. If, for every $a\in I$, we have $f(b)\geq f(a)$ for all $b\in I$ sufficiently close to $a$ with $b>a$, then $f$ is nondecreasing on $I$. If these local inequalities are strict, then $f$ is strictly increasing.} These conclusions give the stated bounds for every $a<a'$ in $I$.
\end{proof}

\subsubsection{Proof of Proposition \ref{own-patience}}\label{appendix-own-patience}

\begin{proof}
    Fix $d_0\in I$, and write $\theta_0=\theta^{d_0}$ and $y_0=y(d_0)$. We suppress $\theta_0$ in the notation for maximizing menus, proposer surpluses, types, and inclusion polytopes. Put $K_0=K(y_0)$, and let $r_0\in R(y_0)$ be the unique supporting aggregate inclusion vector given by \eqref{structural-local-base-aggregate}. As shown at the start of Section \ref{appendix-structural-local-index}, $y_0\in\operatorname{int}Y(\theta_0)$. For $d\in I$, put $y=y(d)$ and write $\gamma=1/d_0-1/d$, which has the same sign as $d-d_0$.

    \proofstep{Restrictions at nearby equilibria.}
    As in the construction at the start of the proof of Lemma \ref{structural-local-index}, choose a neighborhood $V\subset\operatorname{int}Y(\theta_0)$ of $y_0$ such that $S_j(z)>0$ and $\mathcal C_j(z)\subseteq\mathcal C_j(y_0)$ for every $j\in N$ and $z\in V$. Corollary~\ref{equilibrium-payoff-continuity} gives $y\to y_0$ as $d\to d_0$, so $y\in V$ for all sufficiently close $d$. The maximization problem \eqref{problem} depends on $\chi$ and the candidate payoff vector, but not on discount factors. Thus the maximizing menus and proposer surpluses at $y$ under $\theta^d$ coincide with those under $\theta_0$.

    Define $F(y)$ by \eqref{structural-local-aggregate-set}, so it is a non-empty face of $R(y_0)$ by \eqref{structural-local-minimizing-face}. Let $r\in F(y)$ be generated by the equilibrium proposal distributions of the proposers in $K_0$ under $\theta^d$. Proposers outside $K_0$ retain their unique maximizing coalitions. The equilibrium payoff equations are
    \[
        y_j=\delta_j^dp_jS_j(y)+\delta_j^d\bigl[p_{C^j(y_0)}+r^j\bigr]y_j, \qquad j\in N.
    \]
    Since $1/\delta_j^d-1/\delta_j^{d_0}=-\gamma\mathbb I\{j=i\}$, solving for $r^j$ and subtracting \eqref{structural-local-base-aggregate} gives
    \begin{equation}\label{patience-aggregate-change}
        r^j-r_0^j=\frac{p_jS_j(y_0)}{y_{0,j}y_j}(y_j-y_{0,j}) +\frac{p_j}{y_j}\bigl[S_j(y_0)-S_j(y)\bigr] -\gamma\mathbb I\{j=i\},\qquad j\in N.
    \end{equation}
    Since $y\to y_0$ and $\gamma\to0$ as $d\to d_0$, this identity implies $r\to r_0$. The remaining arguments in Step \ref{step-structural-local-faces} of the proof of Lemma \ref{structural-local-index} therefore give, for all sufficiently close $d$,
    \begin{equation}\label{patience-face-directions}
        r-r_0\in L_{\mathcal A(y)},\qquad y-y_0\in L_{\mathcal A(y)}^{\perp}.
    \end{equation}

    \proofstep{Local payoff comparison.}\label{step-patience-local}
    Since $\chi$ is $M^\natural$-concave and $S_j(y)>0$ for every $j\in N$, Lemma \ref{mnatural-tie-structure} gives an equitable grouping $G_1,\ldots,G_\ell$ at $y$. Let $X=[\mathbf1_{G_1}\ \cdots\ \mathbf1_{G_\ell}]$. By part (\ref{maximizing-group-basis}) and \eqref{patience-face-directions}, we have $X^\top(r-r_0)=0$ and $y-y_0=X\xi$ for some $\xi\in\mathbb R^\ell$. Note that the groups, and therefore $X$ and $\xi$, can depend on $d$. If $\ell=0$, then $y=y_0$. Suppose that $\ell>0$, and let $\mathsf C$ be the count matrix at $y$ defined in Lemma \ref{comparative-group-matrix}. The surplus calculation in Step \ref{step-structural-local-groups} of the proof of Lemma \ref{structural-local-index} gives $S_j(y_0)-S_j(y)=(\mathsf C\xi)_t$ for $j\in G_t$. Because the groups are disjoint, $y_j-y_{0,j}=\xi_t$ for $j\in G_t$. Summing \eqref{patience-aggregate-change} over $j\in G_t$ and using $X^\top(r-r_0)=0$ therefore gives
    \begin{equation}\label{patience-group-equations}
        0=\sum_{j\in G_t}(r^j-r_0^j)=A_t\xi_t+B_t(\mathsf C\xi)_t-\gamma\mathbb I\{i\in G_t\}, \qquad t=1,\ldots,\ell,
    \end{equation}
    where
    \[
        A_t=\sum_{j\in G_t}\frac{p_jS_j(y_0)}{y_{0,j}y_j}>0, \qquad B_t=\sum_{j\in G_t}\frac{p_j}{y_j}>0.
    \]
    Let $M=\operatorname{diag}(A_t)_{t=1}^{\ell}+\operatorname{diag}(B_t)_{t=1}^{\ell}\mathsf C$. If $i$ does not belong to any group, \eqref{patience-group-equations} becomes $M\xi=0$. Lemma \ref{comparative-group-matrix} gives $\xi=0$, so $y=y_0$.

    If $i\in G_g$, the same equations give $M\xi=\gamma e_g$, where $e_g$ is the $g$th standard basis vector in $\mathbb R^\ell$. Hence $y_i(d)-y_i(d_0)=\xi_g=\gamma(M^{-1})_{gg}$. By \eqref{comparative-inverse-diagonal-bound}, $(M^{-1})_{gg}>0$, so $y_i(d)-y_i(d_0)$ has the same sign as $\gamma$. Combining this case with the cases in which $y=y_0$, we conclude that, for every sufficiently close $d>d_0$, $y_i(d)\geq y_i(d_0)$, with equality only if $y(d)=y(d_0)$. For sufficiently close $d<d_0$, the inequality is reversed and the same equality conclusion holds.

    \proofstep{Comparison on the interval.}
    Since $d_0$ was arbitrary, the fact in footnote~\ref{fn-local-to-global-monotonicity} implies that $y_i$ is nondecreasing on $I$. Suppose that $d<d'$ and $y_i(d)=y_i(d')$. Monotonicity implies that $y_i$ is constant on $[d,d']$. At each $d_0\in[d,d']$, the local equality conclusion in Step \ref{step-patience-local} therefore implies that $y$ is constant on a neighborhood of $d_0$ in $[d,d']$. Thus $y$ is locally constant on $[d,d']$, and hence constant on this interval. In particular, $y(d)=y(d')$. The converse is immediate by taking the $i$th coordinate of $y$.
\end{proof}

\end{appendices}

\bibliographystyle{aea}
\bibliography{cites}

\clearpage
\pagenumbering{arabic}
\begingroup
\phantomsection
\addcontentsline{toc}{section}{Supplementary Appendix}
\begin{center}
    {\Large Multiplicity and Uniqueness of Stationary Expected Payoffs\\
    in Coalitional Bargaining\par}
    \medskip
    {\large\bfseries Supplementary Appendix\par}
    \medskip
    \large Kirill S. Evdokimov\\
    November, 2026
\end{center}

\setcounter{section}{0}
\setcounter{subsection}{0}
\setcounter{obs}{0}
\setcounter{aprop}{0}
\renewcommand{\thesection}{SA.\arabic{section}}
\renewcommand{\thesubsection}{\thesection.\arabic{subsection}}
\renewcommand{\theobs}{SA.\arabic{obs}}
\renewcommand{\theaprop}{SA.\arabic{aprop}}
\renewcommand{\theHsection}{sa.\arabic{section}}
\renewcommand{\theHsubsection}{\theHsection.\arabic{subsection}}
\renewcommand{\theHobs}{sa.\arabic{obs}}
\renewcommand{\theHaprop}{sa.\arabic{aprop}}

\section{Equilibrium payoffs in Example \ref{ex-delay-multiplicity}}\label{appendix-three-player}\label{appendix-multiplicity-example}

Following Example \ref{ex-delay-multiplicity}, let $n=3$, $p_i=\frac13$, and $\delta_i=\frac{9}{10}$ for all $i\in N$. Let the collection of feasible coalitions be $\mathcal W=\{C\,:\,\lvert C\rvert\geq2\}$ and let $\chi(C)=\min_{i\in C}a_i$ for all $C\in\mathcal W$, where $a_1=\frac25$ and $a_2=a_3=1$.

Let $y$ be an equilibrium payoff vector. When player 1 is recognized, players 2 and 3 together receive at most $y_2+y_3$. When player $i\in\{2,3\}$ is recognized, their combined payoff is at most $1$. If she forms a coalition, its surplus is at most $1$. If she passes, $S_i(y)=0$, so equation \eqref{payoff} implies $y_i=0$, while the other player's continuation payoff is at most $1$. Thus
\[
    y_2+y_3\le \frac9{10}\left(\frac13(y_2+y_3)+\frac23\right)
    \qquad\implies\qquad y_2+y_3\le\frac67<1.
\]
It follows that $S_i(y)\geq \chi(\{2,3\})-\frac67=\frac17>0$ and therefore $y_i>0$ for $i\in\{2,3\}$. In particular, players 2 and 3 never pass.

Now, we can see that the grand coalition is never optimal. Player $1$ strictly prefers $\{1,i\}$ to $\{1,2,3\}$ for $i\in\{2,3\}$. In turn, player $i\in\{2,3\}$ strictly prefers $\{2,3\}$ to $\{1,2,3\}$ for all $y_1\geq0$. The remaining possible maximizing coalitions are:
\[
    \text{Player }1:\,\, C_\varnothing,\,\{1,2\},\,\{1,3\}, \quad \text{Player }2:\,\, \{1,2\},\,\{2,3\}, \quad \text{Player }3:\,\, \{1,3\},\,\{2,3\}.
\]

We next show that $y_2=y_3$. Suppose that $y_2<y_3$. Then, player 1 never proposes $\{1,3\}$, and $S_2(y)\geq S_3(y)$. Let $h_2$ be player 2's probability of proposing $\{1,2\}$, let $h_3$ be player 3's probability of proposing $\{1,3\}$, and let $\mu$ be player 1's probability of forming a non-empty coalition. If $h_3=0$, then trivially $h_2\ge h_3$. If $h_3>0$, then $\{1,3\}$ is optimal for player 3, so $\{1,2\}$ is strictly optimal for player 2 and $h_2=1$, and therefore $h_2\ge h_3$. Then, the equilibrium payoffs of players 2 and 3 satisfy
\begin{align*}
    y_2 & =\frac{3}{10}(S_2(y)+y_2)+\frac{3}{10}y_2+\frac{3}{10}(1-h_3)y_2 \quad\implies\quad y_2=\frac{3S_2(y)}{1+3h_3}\\
    y_3 & =\frac{3}{10}(S_3(y)+y_3)+\frac{3}{10}(1-\mu)y_3+\frac{3}{10}(1-h_2)y_3 \quad\implies\quad y_3=\frac{3S_3(y)}{1+3h_2+3\mu},
\end{align*}
and $h_2\ge h_3$ and $\mu\geq0$ imply $y_2\ge y_3$, a contradiction. The case $y_2>y_3$ is symmetric.

We are now ready to describe all equilibria. Write $y_2=y_3=w$ and note that $y_2+y_3\leq\frac67$ implies $w\leq\frac37$. Therefore, for $i\in\{2,3\}$, we have
\begin{align*}
    & \chi(\{2,3\})-y_2-y_3-\left(\chi(\{1,i\})-y_1-y_i\right)\\
    & \qquad\qquad =1-2w-\left(\frac25-y_1-w\right)=\frac35+y_1-w\ge\frac6{35}+y_1>0,
\end{align*}
so players 2 and 3 always propose $\{2,3\}$ and $S_2(y)=S_3(y)$.

Since players 2 and 3 have the same positive proposer surplus and payoff, their continuation probabilities are equal, so player 1 proposes each of $\{1,2\}$ and $\{1,3\}$ with probability $\frac\mu2$. Then, the equilibrium expected payoffs satisfy
\begin{align*}
    y_1 & =\frac{3}{10}\mu\left(\frac25-w\right)+\frac{3}{10}(1-\mu)y_1,\\
    w & =\frac{3}{10}(1-w)+\frac{3}{10}w+\frac{3}{10}\left(1-\mu+\frac\mu2\right)w \quad\implies\quad w=\frac{6}{14+3\mu}.
\end{align*}

If $\mu=0$, player 1 passes, so $y_1=0$ and $w=\frac37$. If $0<\mu<1$, player 1 is indifferent between passing and proposing $\{1,2\}$ or $\{1,3\}$, so $y_1=0$ and $w=\frac25$, which gives $\mu=\frac13$. Finally, if $\mu=1$, then $w=\frac6{17}$ and $y_1=\frac6{425}$.

Conversely, for $\mu=0,\frac13,1$, player 1's proposer surplus from forming coalition $\{1,2\}$ or $\{1,3\}$ is $-\frac1{35}$, $0$, and $\frac{14}{425}$, respectively. Her prescribed proposal probabilities are therefore optimal. Players 2 and 3 strictly prefer $\{2,3\}$ in each case, and all payoff equations are satisfied. By Proposition~\ref{payoff-correspondence}, these are exactly the equilibrium payoff vectors in this example.

\section{Equilibrium payoffs in Example \ref{ex-no-delay-multiplicity}}\label{appendix-no-delay-multiplicity-payoffs}

Following Example \ref{ex-no-delay-multiplicity}, let $n=3$, $p=\left(\frac{15}{16},\frac1{32},\frac1{32}\right)$, $\delta=\left(\frac{16}{23},\frac{64}{65},\frac{64}{65}\right)$, $\mathcal W=2^N\setminus\{\varnothing\}$, and 
\[
    \chi(\{1\})=1,\quad \chi(\{2\})=\chi(\{3\})=2,\quad \chi(C)=\frac52 \text{ if }|C|=2,\quad \chi(N)=5.
\]

Let $y$ be an equilibrium payoff vector. Since every singleton is feasible, the argument in Section~\ref{sec-no-delay} gives $S_i(y)>0$ for every player. Equation \eqref{payoff} therefore gives $y_i>0$, and no player passes.

We first show that $y_2=y_3$. Suppose that $y_2<y_3$. Replacing player 3 by player 2 in $\{3\}$ or $\{1,3\}$ leaves coalition surplus unchanged and reduces the sum of continuation payoffs. Moreover, $\{2,3\}$ and $N$ are feasible for both $i\in\{2,3\}$. Hence $S_2(y)\geq S_3(y)$.

Let $h_i$ be the probability that player $i\in\{2,3\}$ proposes a coalition containing both players 2 and 3. If $h_2=0$, then trivially $h_2\le h_3$. If $h_2>0$, some such coalition is maximizing for player 2. Its proposer surplus is strictly greater than the proposer surplus from either $\{3\}$ or $\{1,3\}$, so player 3 always recruits player 2 and $h_3=1$. Thus $h_2\leq h_3$. Let $\mu_i$ be the probability that player 1 recruits player $i\in\{2,3\}$. Player 1 strictly prefers $\{1,2\}$ to $\{1,3\}$, while her other feasible coalitions contain both players or neither. Hence $\mu_2\geq\mu_3$.

The payoff equations for players 2 and 3 are
\begin{align*}
    y_2&=\frac2{65}(S_2(y)+y_2)+\frac2{65}h_3y_2+\frac{60}{65}\mu_2y_2
    \quad\implies\quad y_2=\frac{2S_2(y)}{63-2h_3-60\mu_2},\\
    y_3&=\frac2{65}(S_3(y)+y_3)+\frac2{65}h_2y_3+\frac{60}{65}\mu_3y_3
    \quad\implies\quad y_3=\frac{2S_3(y)}{63-2h_2-60\mu_3}.
\end{align*}
Both denominators are positive. Since $S_2(y)\geq S_3(y)$, $h_3\geq h_2$, and $\mu_2\geq\mu_3$, these equations imply $y_2\geq y_3$, a contradiction. The case $y_2>y_3$ is symmetric.

We next determine which coalitions can be proposed. Write $y_2=y_3=w$. Every proposed coalition generates surplus at most $5$, so summing \eqref{payoff} after dividing each player's equation by her discount factor gives
\[
    \frac{y_1}{\delta_1}+\frac{2w}{\delta_2}\leq5.
\]
Since $y_1>0$, it follows that $w<\frac52\delta_2=\frac{32}{13}<\frac52$. The grand coalition is therefore strictly preferred to either $\{1,2\}$ or $\{1,3\}$, because $5-y_1-2w>\frac52-y_1-w$ is equivalent to $w<\frac52$.

If $w\geq2$, then both $2-w$ and $\frac52-2w$ are nonpositive. Players 2 and 3 have strictly positive proposer surplus, so both must propose the grand coalition. Their common proposer surplus is then $S_2(y)=S_3(y)=5-y_1-2w\leq1-y_1<1$. Player $i\in\{2,3\}$ receives $S_i(y)+w$ when recognized and at most $w$ otherwise. Her payoff equation therefore gives
\[
    w\leq\frac{64}{65}\left(w+\frac{S_i(y)}{32}\right)
    \qquad\implies\qquad w\leq2S_i(y)<2,
\]
a contradiction. Thus $w<2$, and player 1 strictly prefers the grand coalition to her singleton because $5-y_1-2w>1-y_1$ is equivalent to $w<2$.

Player $i\in\{2,3\}$ receives $w$ whenever player 1 is recognized and can obtain an allocation of $2$ by forming her singleton when recognized herself. Since all other payoff terms are nonnegative, her payoff equation gives
\[
    w\geq\frac{64}{65}\left(\frac{15}{16}w+\frac2{32}\right)
    \qquad\implies\qquad w\geq\frac45>\frac12.
\]
For $i\in\{2,3\}$, the singleton coalition therefore gives strictly greater proposer surplus than coalition $\{2,3\}$ because $2-w>\frac52-2w$ is equivalent to $w-\frac12>0$. The remaining possible maximizing coalitions are:
\[
    \text{Player }1:\,\, \{1,2,3\}, \qquad \text{Player }2:\,\, \{2\},\,\{1,2,3\}, \qquad \text{Player }3:\,\, \{3\},\,\{1,2,3\}.
\]

For $i\in\{2,3\}$, $h_i$ is therefore the probability that player $i$ proposes the grand coalition. Since players 2 and 3 have the same positive proposer surplus and payoff, their continuation probabilities are equal. Hence
\[
    p_1+p_2+p_3h_3=p_1+p_3+p_2h_2
    \qquad\implies\qquad h_2=h_3=h.
\]

Now we can write the payoff equations as follows:
\begin{align*}
    y_1&=\frac{15}{23}(5-2w)+\frac{h}{23}y_1,\\
    w&=\frac2{65}\bigl[(1-h)2+h(5-y_1-w)\bigr]
       +\frac{60}{65}w+\frac{2h}{65}w.
\end{align*}
For $h\in[0,1]$, these equations have the unique solution
\[
    y_1(h)=\frac{51-36h}{23-13h},
    \qquad
    w(h)=\frac{92-16h-6h^2}{5(23-13h)}.
\]
For players 2 and 3, the grand coalition's advantage in proposer surplus over the singleton is
\[
    \Delta(h)=3-y_1(h)-w(h)
    =\frac{(2h-1)(3h+2)}{5(23-13h)}.
\]
If $0<h<1$, both coalitions are proposed with positive probability, so optimality requires $\Delta(h)=0$ and therefore $h=\frac12$. The remaining possibilities are $h=0$ and $h=1$. Thus every equilibrium payoff vector must be one of
\[
    y(0)=\left(\frac{51}{23},\frac45,\frac45\right),
    \qquad y\!\left(\frac12\right)=(2,1,1),
    \qquad y(1)=\left(\frac32,\frac75,\frac75\right).
\]

Conversely, each of these vectors satisfies $\frac12<w<2$, so the strict comparisons excluding pairs and player 1's singleton hold. For $h=0,\frac12,1$, the proposer surplus from the grand coalition is $\frac{136}{115}$, $1$, and $\frac7{10}$, respectively, so passing is strictly suboptimal for every player. For players 2 and 3, the difference in proposer surplus $\Delta(h)$ is $-\frac2{115}$, $0$, and $\frac1{10}$, respectively. Their prescribed proposal probabilities are therefore optimal, and all payoff equations are satisfied. By Proposition~\ref{payoff-correspondence}, these are exactly the equilibrium payoff vectors in this example.

At each of $y(0)$ and $y(1)$, each proposer has a unique maximizing coalition and the maximizing coalitions remain unchanged for nearby candidate payoff vectors and primitives. Therefore, the payoff system \eqref{payoff-interior} is locally smooth and has a nonsingular Jacobian with respect to $y$ by Lemma~\ref{notie-local-degree}. Then, for every sufficiently nearby $(\chi,p,\delta)$, the implicit function theorem gives two distinct no-tie equilibrium payoff vectors, one near $y(0)$ and one near $y(1)$. Moreover, each has local index $+1$ by Lemma~\ref{notie-local-degree}, and therefore Additivity and Lemma~\ref{degree-one} guarantee at least one additional equilibrium payoff vector.

We can also verify the feedback mechanism described after Example~\ref{ex-no-delay-multiplicity}. Varying the common proposal probability $h$ around $\frac12$ gives
\[
    y_1'\!\left(\frac12\right)=-\frac{20}{33},
    \qquad w'\!\left(\frac12\right)=\frac{86}{165},
    \qquad \Delta'\!\left(\frac12\right)=\frac{14}{165}>0.
\]
Thus players 2 and 3 become more expensive for proposer 1 to recruit, but player 1's continuation payoff falls by more than either other player's continuation payoff rises. The combined cost of recruiting the other two players falls for each of proposers 2 and 3, making the grand coalition more attractive relative to the singleton and reinforcing the increase in $h$.

\section{Limited harm and no delay}\label{appendix-limited-harm}

Observation~\ref{negative-externality-no-delay} in Section \ref{multiplicity} suggests a recursive restriction on the primitives to verify Assumption ND. If $y$ is an equilibrium payoff vector and $S_i(y;\theta)=0$, then $y_i=0$. Moreover, no coalition with strictly positive proposer surplus can include player $i$, because she could then obtain strictly positive proposer surplus by proposing that coalition herself. Thus any delay equilibrium partitions $N$ into a set $A$ of players who form strictly profitable coalitions among players in $A$ and a \textit{fringe} set $N\setminus A$ of harmful players who have zero payoffs.

The restriction, called \textit{limited harm}, is formulated using a family of reduced games. For any non-empty subset $A\subseteq N$, let $\mathcal W_A=\{C\in\mathcal W:C\subseteq A\}$ and $\mathcal W_A^i=\{C\in\mathcal W_A:i\in C\}$. Write $p_A=\sum_{i\in A}p_i$. For each $\lambda\in[0,1-p_A]^A$, consider the bargaining game with player set $A$, feasible coalitions $\mathcal W_A$, the restriction of $\chi$ to $\mathcal W_A$, recognition probabilities $p_i/p_A$, and discount factors
\[
    \delta_i^A(\lambda)=\frac{\delta_i p_A}{1-\delta_i\lambda_i},\qquad i\in A.
\]
These discount factors belong to $(0,1)$ because $\delta_i(p_A+\lambda_i)\leq\delta_i<1$. The bargaining protocol is otherwise unchanged. We allow $A$ to be a singleton and $\mathcal W_A^i$ to be empty. A player with $\mathcal W_A^i=\varnothing$ can only pass.

The parameter $\lambda_i$ is intended to represent the contribution of proposers outside $A$ to player $i$'s continuation probability in the original game, through passing or forming a coalition containing $i$. The discount factors $\delta_i^A(\lambda)$ incorporate these contributions into the payoff equations while recognition probabilities are normalized to $p_i/p_A$. Below we restrict $\lambda$ to contributions generated by proposal distributions of fringe players.

For $z\in\mathbb R_+^A$ and each $i\in A$, define
\[
    \begin{aligned}
        S_i^A(z)&=\max_{C\in\mathcal W_A^i\cup\{C_\varnothing\}}(\chi(C)-z_C)\qquad\text{and}\\
        \mathcal C_i^A(z)&=\argmax_{C\in\mathcal W_A^i\cup\{C_\varnothing\}}(\chi(C)-z_C).
    \end{aligned}
\]
Let $R_A(z)$ and $C_A^i(z)$ be the analogues of $R(y)$ and $C^i(y)$ with proposer set $A$, maximizing menus $\mathcal C_j^A(z)$, and the original recognition probabilities $p_j$. By the same argument as in the proof of Proposition~\ref{payoff-correspondence}, for fixed $\lambda\in[0,1-p_A]^A$, a vector $z\in\mathbb R_+^A$ is an equilibrium payoff vector of the reduced game if and only if there exists $r\in R_A(z)$ such that
\[
    z_i=\delta_i^A(\lambda)\left[\frac{p_i}{p_A}S_i^A(z)+\frac{p_{C_A^i(z)}+r^i}{p_A}z_i\right],\qquad i\in A.
\]
Substituting the definition of $\delta_i^A(\lambda)$ and rearranging gives
\begin{equation}\label{limited-harm-reduced-payoffs}
    z_i=\frac{\delta_ip_iS_i^A(z)}{1-\delta_ip_{C_A^i(z)}-\delta_ir^i-\delta_i\lambda_i},\qquad i\in A.
\end{equation}

To relate these reduced games to the original game, extend $z\in\mathbb R_+^A$ to $z^A\in\mathbb R_+^n$ by setting $z_i^A=z_i$ for $i\in A$ and $z_i^A=0$ otherwise. For each fringe player $j\notin A$, we allow mixing over the set $\mathcal D_j^A(z)=\{C\in\mathcal W^j\cup\{C_\varnothing\}:\chi(C)-z_C^A=0\}$ of coalitions yielding zero proposer surplus. Passing moves the game to the next period and gives every player in $A$ her continuation payoff. Forming a non-empty coalition ends the game and gives continuation payoffs only to the members it includes. Let $\mathcal H_A(z)\subseteq[0,1-p_A]^A$ collect the contributions of fringe players to continuation probabilities:
\[
    \mathcal H_A(z)=\left\{
        \lambda\in[0,1-p_A]^A:\lambda_i=\sum_{j\notin A}p_j\sum_{\substack{C\in\mathcal D_j^A(z):\\ C\ni i\text{ or }C=C_\varnothing}}\rho_j(C),\ \rho_j\in\Delta(\mathcal D_j^A(z))
    \right\}.
\]

I will say that $z$ is an \textit{augmented reduced equilibrium} on $A$ if it is an equilibrium payoff vector of the reduced game for some $\lambda\in\mathcal H_A(z)$. Let $\mathcal E_A(\theta)$ denote the set of such payoff vectors. This definition does not require the zero-surplus proposals of fringe players to be optimal in the original game.

For a fringe player $i\notin A$, define her \textit{entry surplus} against such a payoff vector $z$ by $m_{i\mid A}(z)=\max_{C\in\mathcal W^i}\bigl(\chi(C)-z_C^A\bigr)$. This is the largest net surplus that player $i$ can obtain by forming a non-empty coalition in the original game when the players in $A$ are priced at $z$ and the fringe players are priced at zero.

\begin{assumption}{LH}
    For every proper non-empty subset $A\subset N$ and every augmented reduced equilibrium payoff vector $z$ on $A$, there exists at least one fringe player $i\notin A$ such that ${m_{i\mid A}(z)>0}$.
\end{assumption}

Assumption LH rules out delay configurations in which some proper subgroup sustains its own continuation payoffs while no fringe player can form a profitable coalition. The next proposition shows that it characterizes Assumption ND exactly.

\begin{aprop}\label{lh-no-delay}
    Fix $\theta\in\Theta$. Assumption LH holds if and only if $\theta\in\Theta^{\mathrm{ND}}$.
\end{aprop}

\begin{proof}
    Suppose first that Assumption LH holds. Let $y\in\mathcal E(\theta)$, and let $(\rho_j)_{j\in N}$ be the proposal distributions from \eqref{payoff}. Suppose, toward a contradiction, that $S_j(y)=0$ for some player $j\in N$. Let $A=\{i\in N:S_i(y)>0\}$. Then $A$ is a proper subset of $N$. Moreover, $A$ is non-empty by the same argument as in the proof of Proposition \ref{monotone-no-delay}.

    For every fringe player $j\notin A$, we have $S_j(y)=0$, so equation \eqref{payoff} gives $y_j=0$. Let $z=(y_i)_{i\in A}\in\mathbb R_+^A$. Then $y=z^A$, so every $C\in\mathcal C_j(y)$ for $j\notin A$ satisfies $\chi(C)-z_C^A=\chi(C)-y_C=S_j(y)=0$. Hence $\mathcal C_j(y)\subseteq\mathcal D_j^A(z)$ for every $j\notin A$. Define
    \[
        \lambda_i=\sum_{j\notin A}p_j\sum_{\substack{C\in\mathcal C_j(y):\\C\ni i\text{ or }C=C_\varnothing}}\rho_j(C),\qquad i\in A.
    \]
    Then $\lambda\in\mathcal H_A(z)$.

    For every $i\in A$ and $C\in\mathcal C_i(y)$, we have $C\subseteq A$. Otherwise, some $j\in C\setminus A$ could form the same coalition $C$, giving $S_j(y)\geq\chi(C)-y_C=S_i(y)>0$, contradicting $j\notin A$. Thus every coalition in $\mathcal C_i(y)$ belongs to $\mathcal W_A^i$. It follows that $S_i^A(z)=S_i(y)$ and $\mathcal C_i^A(z)=\mathcal C_i(y)$ for every $i\in A$. The distributions $(\rho_j)_{j\in A}$ therefore induce $r_A\in R_A(z)$. For every $i\in A$, equation \eqref{payoff} gives 
    \[
        y_i=\frac{\delta_ip_iS_i(y)}{1-\delta_ip_{C_A^i(z)}-\delta_ir_A^i-\delta_i\lambda_i}=\frac{\delta_ip_iS_i^A(z)}{1-\delta_ip_{C_A^i(z)}-\delta_ir_A^i-\delta_i\lambda_i}.
    \]
    Thus $z$ satisfies \eqref{limited-harm-reduced-payoffs} with $r=r_A$, and $\lambda\in\mathcal H_A(z)$ implies $z\in\mathcal E_A(\theta)$.

    Assumption LH implies that there exists some $j\notin A$ such that $m_{j\mid A}(z)>0$. But since $y=z^A$ and $S_j(y)=0$, we have $m_{j\mid A}(z)=\max_{C\in\mathcal W^j}\bigl(\chi(C)-z_C^A\bigr)=\max_{C\in\mathcal W^j}\bigl(\chi(C)-y_C\bigr)\le 0$, a contradiction. Therefore $S_i(y)>0$ for all $i\in N$ and $\theta\in\Theta^{\mathrm{ND}}$.

    For the converse, suppose that Assumption LH fails. Then there are a proper non-empty subset $A\subset N$ and $z\in\mathcal E_A(\theta)$ such that $m_{j\mid A}(z)\leq0$ for every $j\notin A$. Set $y=z^A$. Choose $\lambda\in\mathcal H_A(z)$ such that $z$ is an equilibrium payoff vector of the corresponding reduced game. Let $\rho_j\in\Delta(\mathcal C_j^A(z))$, $j\in A$, be the proposal distributions in such an equilibrium, and choose $\rho_j\in\Delta(\mathcal D_j^A(z))$, $j\notin A$, that generate $\lambda$.

    Every coalition $C$ containing a fringe player $j$ satisfies $\chi(C)-y_C\leq m_{j\mid A}(z)\leq0$. Therefore, we have
    \[
        \begin{aligned}
            S_i(y)&=S_i^A(z), & \mathcal C_i^A(z)&\subseteq\mathcal C_i(y), && i\in A,\\
            S_j(y)&=0, & \mathcal D_j^A(z)&=\mathcal C_j(y), && j\notin A.
        \end{aligned}
    \]
    Thus $\rho_j\in\Delta(\mathcal C_j(y))$ for every $j\in N$. For $y$ and $(\rho_j)_{j\in N}$, the payoff equations \eqref{payoff} in the original game hold for $i\in A$ by \eqref{limited-harm-reduced-payoffs}, and for $j\notin A$ because $y_j=S_j(y)=0$. Hence Proposition~\ref{payoff-correspondence} gives $y\in\mathcal E(\theta)$, so $\theta\notin\Theta^{\mathrm{ND}}$. This proves the converse by contraposition.
\end{proof}

Checking Assumption LH can be demanding because it involves the augmented reduced equilibria of every proper non-empty group of players. The condition can nevertheless be substantially less restrictive than the condition in Proposition~\ref{max-value-no-delay}. In the setting of Observation~\ref{negative-externality-no-delay}, the calculation in Section~\ref{appendix-limited-harm-example} shows that LH holds if and only if $a_1>\frac37$, compared with the bound $a_1>\frac67$ from Proposition \ref{max-value-no-delay}.

\subsection{Limited harm in Example~\ref{ex-delay-multiplicity}}\label{appendix-limited-harm-example}

\begin{obs}\label{negative-externality-limited-harm}
    Fix the primitives of Example~\ref{ex-delay-multiplicity}, except let $a_1\in(0,1)$ vary while $a_2=a_3=1$. Assumption LH holds if and only if $a_1>\frac37$.
\end{obs}

\begin{proof}
    We first show that $a_1>\frac37$ is sufficient. Consider any proper non-empty subset $A\subset N$ and any augmented reduced equilibrium $z$ on $A$. If $A$ is a singleton, no coalition is feasible in the reduced game, so $z=0$ and every fringe player has positive entry surplus. If $A$ is a pair, its only feasible coalition must have positive proposer surplus. Otherwise \eqref{limited-harm-reduced-payoffs} would give $z=0$, contradicting $\chi(A)>0$. Hence $z_A<\chi(A)$, and both players always propose $A$.

    If $A=\{1,2\}$, then $z_1+z_2<a_1<1$, so fringe player 3 can propose $\{2,3\}$ and obtain positive proposer surplus $1-z_2$. The case $A=\{1,3\}$ is symmetric.

    Let $A=\{2,3\}$. Players 2 and 3 always propose $A$. With the original recognition probabilities, these proposals contribute $p_A=\frac23$ to each player's continuation probability. Since for each $i\in\{2,3\}$ we have $\lambda_i\leq p_1=\frac13$, \eqref{limited-harm-reduced-payoffs} gives
    \[
        z_i=\frac{\frac3{10}(1-z_2-z_3)}{1-\frac9{10}(\frac23+\lambda_i)}\leq3(1-z_2-z_3), \qquad i\in\{2,3\}.
    \]
    Summing yields $z_2+z_3\leq\frac67$, and therefore
    \[
        m_{1\mid A}(z)=\max\{a_1-z_2,a_1-z_3,a_1-z_2-z_3\}=a_1-\min\{z_2,z_3\}\geq a_1-\frac37>0.
    \]
    Thus LH holds whenever $a_1>\frac37$.

    To show necessity, let $a_1\leq\frac37$ and consider the reduced game on $A=\{2,3\}$ with $\lambda_2=\lambda_3=\frac13$. If both players propose $A$, \eqref{limited-harm-reduced-payoffs} becomes
    \[
        z_i=3(1-z_2-z_3),\qquad i\in\{2,3\}, \qquad\implies\qquad z_2=z_3=\frac37.
    \]
    At this vector, proposing $A$ gives players 2 and 3 proposer surplus $\frac17>0$, so $z$ is an equilibrium payoff vector of the reduced game. In the original game, passing by player 1 generates this $\lambda$, so $\lambda\in\mathcal H_A(z)$. Thus $z$ is an augmented reduced equilibrium. The only fringe player is player 1, whose entry surplus is $m_{1\mid A}(z)=a_1-\frac37\leq0$. Hence LH fails for $a_1\leq\frac37$.
\end{proof}

\endgroup

\end{document}